\pdfoutput=1
\documentclass[%
onecolumn,
preprint,
amsmath,amssymb,
aps,
prfluids,
]{revtex4-2}
\usepackage{graphicx}
\usepackage{dcolumn}
\usepackage{bm}

\usepackage{indentfirst}
\usepackage{amsmath}
\usepackage{amssymb}
\usepackage{soul}
\usepackage{tikz}
\newcommand*\circled[1]{\tikz[baseline=(char.base)]{
  \node[shape=circle,draw,inner sep=1pt] (char) {#1};}}
 \usepackage{setspace}
 \usepackage{csvsimple}
\begin{document}

\preprint{APS/123-QED}

\title{Drag force on spherical particles trapped at a liquid interface}
\author{Zhi Zhou}
\author{Petia M. Vlahovska}%
\email{petia.vlahovska@northwestern.edu}
\author{Michael J. Miksis}%
 \email{miksis@northwestern.edu}
\affiliation{%
Department of Engineering Sciences and Applied Mathematics, Northwestern University\\
Evanston, IL 60208 
}%




\date{\today}

\begin{abstract}
The dynamics of particles attached to an interface separating two immiscible fluids are encountered in a wide variety of applications. Here we present a combined asymptotic and numerical investigation of the fluid motion past spherical particles attached to a deformable interface undergoing uniform creeping flows in the limit of small Capillary number and small deviation of the contact angle from $90^\circ$.  Under the assumption of a constant three-phase contact angle, we calculate the interfacial deformation around an isolated particle and a particle pair. Applying the Lorentz reciprocal theorem to the zeroth-order approximation corresponding to spherical particles at a flat interface and  the first correction in Capillary number and correction contact angle allows us to obtain explicit analytical expressions for the hydrodynamic drag in terms of the zeroth-order approximations and the correction deformations. The drag coefficients are computed as a function of the three-phase contact angle, the viscosity ratio of the two fluids, the Bond number, and the separation distance between the particles. In addition, the capillary force acting on the particles due to the interfacial deformation is calculated.
    
\end{abstract}

\maketitle


\section{Introduction}
The problem of particles attached to an interface between two immiscible fluids has been extensively studied for many decades and has a wide variety of engineering and medical applications, such as the formations of Pickering emulsions and particle monolayers. 
There exists a number of excellent texts and review articles focused on this topic \cite{Leal1980,Maldarelli2022,binks2006}. 
The simplest problem that has been studied consists of a single rigid sphere translating along a flat fluid interface between two immersible fluids at low Reynolds number. 
Having a better estimate of the drag on this particle would have an impact on many applications such as modeling the collective motion of particles on a drop in an applied electric field \cite{Yi2021}. 
The hydrodynamic drag force exerted on a sphere can be written as $F_D = -6\pi \mu_1 a U f(\mu_2/\mu_1,b/a,\Delta \rho)$, where $f$ is a dimensionless drag coefficient, $a$ is the radius of the sphere, $b$ is the immersion depth into the upper fluid, $U$ is the translational velocity, and $\mu_1$ and $\mu_2$ are the viscosities of the two fluid phases (see Fig. \ref{fig:single-setup12}(a) for a sketch of the problem). For the simple case of a homogeneous fluid, $f = 1.$  A variety of experimental and theoretical studies have obtained the drag coefficient in terms of the immersion depth and the viscosity ratio. The analytical solution for the drag force acting on two fused spheres obtained by Zabarankin \cite{Zabarankin2007} provides the solution for a translating sphere at a flat gas/liquid interface with immersion depth $b$ and for $0 < \theta < \pi/2$ \cite{Dani2015,dorr2016}. The calculation was extended by D\"orr et al.  \cite{dorr2016} to cover the full range of contact angles. Dani et al. \cite{Dani2015} and D\"orr et al. \cite{dorr2016} assumed that the three-phase contact line (TCL) is pinned to the particle surface, which prevents particle rotation. D\"orr and Hardt \cite{dorr2015} considered particle rotation in their problem by allowing it to rotate until the hydrodynamic torque and the torque caused by the interfacial tension are balanced, and the steady-state interfacial deformation was calculated \cite{dorr2015}.

Numerical calculations of the drag coefficient $f$ were carried out by \cite{Danov1995,Danov2000,Das2018,Pozrikidis2007}. Danov et al. and Das et al. obtained the drag force acting on a sphere straddling a flat gas-liquid interface  \cite{Danov1995, Danov2000} and a spherical interface at an arbitrary viscosity ratio \cite{Das2018}. Pozrikidis \cite{Pozrikidis2007} solved the problem of a spherical particle at an interface in the presence of a simple shear flow centered at the sphere. As the immersion depth into the liquid phase increases, the drag coefficient $f$ is found to increase monotonically. 
A more recent numerical study by Loudet et al. \cite{Loudet2020} calculated the two-dimensional drag on a circular cylinder straddling a deformable fluid interface at an arbitrary viscosity using a phase-field model. Instead of pinning the TCL, the three-phase contact angle was prescribed.  Hemauer et al. \cite{Hemauer2021} further extended Loudet et al. \cite{Loudet2020}'s work by including particle rotation.

The pair interaction of particles at low Reynolds number in a homogeneous fluid has been well addressed in the literature. 
For particles attached to a fluid interface, capillary interactions arise from the interfacial deformation around them \cite{Danov2010,Kralchevsky2000}. D\"orr and Hardt \cite{dorr2015} examined the capillary interaction between two particles via the linear superposition of the single-particle interfacial deformation, assuming a large particle separation. The capillary interaction between two mutually approaching particles was studied by Dani et al. \cite{Dani2015}, where the viscous drag due to the mutual approach is approximated by multiplying the single-particle drag by the mobility function, which accounts for the increased hydrodynamic resistance as the particles travel closer to each other.

In this work, we wish to further understand the influence of interfacial deformations on the drag force acting on particles at an interface.
The problem is in general complex because of the coupling of fluid flow, interfacial deformation and contact line dynamics. 
Hence a fully numerical approach would be complex. Here we developed an asymptotic solution approach which allows for a direct and straightforward examination of the impact of the physical parameters on the flow. It has been noted in the literature \cite{Maldarelli2022} that uniform flow past a sphere can represent flow past a sphere at an interface in the limit of small Capillary number. This solution is valid for arbitrary viscosity ratio of upper and lower fluids as long as the interface intersects the particle at its equator at $90^\circ$. This observation allows for an estimate of the Stokes drag where the viscosity is the average of the two phases. Here we use this result as the leading order solution in a perturbation analysis for small Ca and small deviation of the contact angle from $90^\circ$. We obtain the analytical solution for the interfacial deformation around a single particle. In the two-particle deformation case, a straightforward numerical solution approach is developed. We apply the Lorentz reciprocal theorem to the zeroth-order approximations for a spherical particle at a flat interface and the first corrections to obtain explicit analytical expressions for the hydrodynamic drag. We compute the drag force as a function of the three-phase contact angle, the viscosity ratio between the two fluids, the separation distance between the particles, and the Bond number (a dimensionless number that measures the relative importance of gravity and surface tension forces).
\section{Fluid motion past a single particle}
\label{sect:single-particle}
\begin{figure}
    \centering
    \includegraphics[width=16cm]{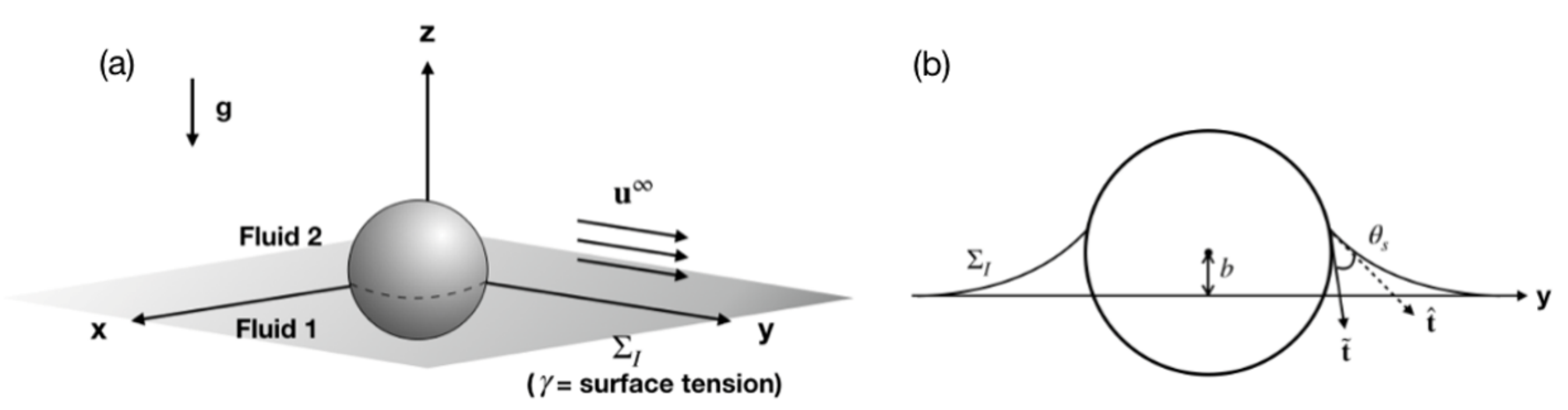}
    \caption{Illustrations of (a) a spherical particle on an flat interface between two immiscible viscous fluids and (b) the cross-section view of a spherical particle at a deformable interface with contact angle $\theta_s$ and immersion depth $b$.}
    \label{fig:single-setup12}
\end{figure}

Consider the uniform creeping flow past a fixed spherical particle of radius $a$ located approximately midway at an interface between two viscous fluids. The lower fluid phase is denoted by $i =1$ and the upper fluid phase by $i = 2$ (see Fig. \ref{fig:single-setup12}(a)). The fluid viscosities and densities are denoted by $\mu_i$ and $\rho_i$, respectively.  
We assume the fluid interface is deformable and the deformation remains small, which requires that the surface tension forces are large relative to the viscous forces, i.e., the Capillary number, Ca $=\mu_1 U/\gamma$, is small. Here, $U$ is the absolute value of the uniform background flow velocity, and $\gamma$ is the interfacial tension.  At the TCL, we enforce a constant contact angle $\theta_s$ on the particle surface, where the assumption of small interfacial deformation requires $\theta_s$ to be close to $90^\circ$. Note that if the top and bottom fluids were the same and $\theta_s = 90^\circ$, this would represent uniform flow past a sphere in a homogeneous flow with its TCL located at the equator of the sphere. 
In addition, we allow for the particle to be displaced vertically and set $\mathbf{x}_P = b \hat{\mathbf{e}}_z$ as the position of the particle's center of mass. Particle rotation is ignored.

The motion of the fluids is governed by the Stokes equations:
\begin{align}
    -\nabla p^{i} + \lambda_i \nabla^2 \mathbf{u}^{i} = &  \mathbf{0} \quad \mbox{ in fluid }i, \label{eqn:single-dimensionless-mom1}\\ 
     \nabla\cdot \mathbf{u}^{i} = &  0 \quad \mbox{ in fluid }i, \label{eqn:single-dimensionless-cont}
\end{align}
where $p^i$, $\mathbf{u}^i$, and $
\lambda_i$ are the respective pressure, velocity, and viscosity in fluid phase $i$ ($i = 1, 2$), where $\lambda_1 = 1$ and $\lambda_2 = \lambda = \mu_2/\mu_1$.  
All variables are made dimensionless using the characteristic unit of length $a$, the characteristic unit of velocity $U$, the characteristic unit of pressure $\mu_1 U/a$, and the viscosity in fluid 1, $\mu_1.$ 

Let $\Sigma_{P_{i}}$ denote the particle surface immersed in fluid $i$, and $\Sigma_I = \{ (x,y,z) \vert F_s(x,y,z) = z - h(x,y) = 0 \}$ denote the fluid interface.
Along the fluid interface $\Sigma_I$, the normal velocity vanishes, the tangential velocity and shear stress are continuous, and the normal stress is discontinuous, i.e.,
\begin{align}
& \mathbf{u}^i \cdot \hat{\mathbf{n}}=0, \label{eqn:single-zero-normal-vel}\\
  &   \hat{\mathbf{t}}\cdot (\mathbf{u}^2 - \mathbf{u}^1) = 0, \label{eqn:single-cont-tang-vel}\\
   &      \hat{\mathbf{t}}\cdot (\bm{\sigma}^2 - \bm{\sigma}^1) \cdot \hat{\mathbf{n} } = 0,\label{eqn:single-tang-stress-cond}\\
   &     (\bm{\sigma}^2 - \bm{\sigma}^1) \cdot \hat{\mathbf{n} } =  \frac{1}{\mbox{Ca}}(\nabla\cdot\hat{\mathbf{n}})\hat{\mathbf{n}} + \frac{\mbox{Bo}}{\mbox{Ca}} h \hat{\mathbf{n}}, \label{eqn:single-stressJumpCond1}
\end{align}
where $\bm{\sigma}^i = -p \mathbf{I} + \lambda_i \left(\nabla \mathbf{u}^i + (\nabla \mathbf{u}^i) ^T\right)$ is the stress tensor, $\hat{\mathbf{n}}$ and $\hat{\mathbf{t}}$ are the unit normal and tangential vectors , respectively, to the fluid interface $\Sigma_I$. The Capillary number, Ca, and the Bond number, Bo, are dimensionless parameters defined as
$$ \mbox{Ca}=\frac{\mu_1 U}{\gamma} \quad \mbox{ and } \mbox{Bo} = \frac{(\rho_1 - \rho_2) g a^2 }{\gamma},$$
where $g$ is the acceleration of gravity. 
At the particle surface $\Sigma_{P_i}$, we impose the no-slip and no-penetration conditions: 
\begin{align}
   &  \Tilde{\mathbf{n}}\cdot \mathbf{u}^i  = 0 , \label{eqn:single-noslip-normal}\\
    & \Tilde{\mathbf{t}}\cdot \mathbf{u}^i = 0, \label{eqn:single-noslip-tang}
\end{align}
where $\Tilde{\mathbf{n}}$ and $\Tilde{\mathbf{t}}$ denote the unit normal and tangential vectors, respectively, to the particle surface $\Sigma_{P_i}$.
Far from the particle, the velocity field approaches the uniform background flow: 
\begin{align}
   \mathbf{u}^i \rightarrow \mathbf{u}^\infty = \hat{\mathbf{e}}_y, \label{eqn:single-far-field-cond}
\end{align}

At the TLC, we define the contact angle $\theta_s$ to be the angle between the tangent to the particle surface $\Sigma_{P_i}$ and the tangent to the interface $\Sigma_I$, both in the plane containing the normal to the TCL, which is illustrated in Fig. \ref{fig:single-setup12}(b). The constant contact angle condition is given by 
\begin{align}
        \Tilde{\mathbf{n}}\cdot \hat{\mathbf{n}} = \cos{\theta_s}.\label{eqn:single-contactAngleCond1}
    \end{align}

\
\subsection{Asymptotic expansions}
Assume $$\mathrm{Ca} \ll 1, \quad \theta_s = \pi/2+\delta \tilde{\theta_s} , \quad \mbox{ with }   \delta \ll 1,\mbox{ } \tilde{\theta}_s = \mathcal{O}(1), \quad \mbox{ and } \mbox{Bo} = \mathcal{O}(1), $$ 
where $\delta$ is the small parameter that describes the scale of the contact angle's deviation from $90^\circ. $
We consider the following two-parameter asymptotic expansion for any quantity $f$:
\begin{align}
     f=  f^{(0,0)}+ & \mbox{Ca} f^{(1,0)} +  \delta f^{(0,1)}  + \cdots, \label{eqn:single-asmExpansions-u} 
\end{align}
and for convenience, we omit the superscript $i$ when referring to quantities in both fluids. Although we will only consider the leading order behaviors, we introduce an expansion in Ca and $\delta$ so that the origin of the resulting forces is clear.
  Here $\mathbf{u}^{(0,0)}$ is given by 
  \begin{align}
      \mathbf{u}^{(0,0)} =\frac{1}{4\rho^5} \left( - 3xy(\rho^2-1)\hat{\mathbf{e}}_x + (-3y^2(\rho^2-1) + (\rho-1)( 4\rho^2+\rho+1 )\rho^2 ) \hat{\mathbf{e}}_y - 3yz(\rho^2-1) \hat{\mathbf{e}}_z  \right), \label{eqn:single-leading-sol-u}
  \end{align}
 and, to within an arbitrary additive constant,
  \begin{align}
      p^{(0,0)} = \left\{\begin{array}{ll}
          - \lambda 3 y /(2\rho^3) &  z> 0  \\
            -  3 y /(2\rho^3) &  z < 0 
      \end{array} \right.,  \label{eqn:single-leading-sol-p}
  \end{align}
  where $\rho = \sqrt{x^2 + y^2 + z^2}$ (\cite{leal2010}). 
  Note that   
  the velocity \eqref{eqn:single-leading-sol-u} satisfies the no-slip conditions \eqref{eqn:single-noslip-normal} and \eqref{eqn:single-noslip-tang}, the far field condition \eqref{eqn:single-far-field-cond}, and the velocity conditions \eqref{eqn:single-zero-normal-vel} and \eqref{eqn:single-cont-tang-vel}. The tangential stress along $z = 0$ is zero so Eq. \eqref{eqn:single-tang-stress-cond} is satisfied. The normal stress condition is not identically satisfied except in the Ca $\rightarrow 0$ limit of a flat interface.
  
  The effect of a particle density is to raise or lower the center of mass of the particle. This can be accounted for by perturbing the particle position from the origin, i.e., 
  \begin{align}
      \mathbf{x}_P = \delta \Tilde{b} \hat{\mathbf{e}}_z.\label{eqn:single-asmExpansions-b}
  \end{align} 
 The actual impact of a specific particle density can then be predicted afterwards by a balance of vertical forces. For simplicity, we scale the immersion depth $b$ as $\delta \tilde{b}$ and assume the effect of the background flow on the particle position is accounted for by the value $\tilde b$. The interface shape $h$ is perturbed from the flat interface, i.e., 
  \begin{align}
      h= &  \mbox{Ca}h^{(1,0)} +  \delta h^{(0,1)}  + \cdots,\label{eqn:single-asmExpansions-h}
  \end{align}
  where $\mbox{Ca}h^{(1,0)}$ describes the interfacial deformation induced by the background flow $\mathbf{u}^\infty $, and $\delta h^{(0,1)}$ is the static deformation induced by the contact angle, which describes the equilibrium interface shape in the absence of flow.
 Note that for the two-parameter asymptotic solutions to be valid, we require $\delta >\mbox{Ca}^2$ and 
$\delta^2 < \mbox{Ca}$, which implies $ \delta \sim \mbox{Ca}^\alpha$ with $\frac{1}{2} < \alpha <2. $ Also note that this separation of scales allows us to independently consider the impacts of immersion depth and capillarity. 
  
Substituting Eqs. \eqref{eqn:single-asmExpansions-u} - \eqref{eqn:single-asmExpansions-h} into Eqs. \eqref{eqn:single-dimensionless-mom1} - \eqref{eqn:single-dimensionless-cont} along with the boundary conditions on the particle surface and the fluid interface,
\begin{align}
    & \Sigma_{P_i} = \{ (x,y,z)\vert x^2 + y^2 + (z-\delta \Tilde{b})^2 = 1 \}, \\
    & \Sigma_{I} = \{ (x,y,z) \vert z = \mbox{Ca}h^{(1,0)} + \delta h^{(0,1)} \},
\end{align}
we can now collect terms with similar powers of Ca and $\delta$. This result is discussed below. Additional details can be found in \cite{zhou2022}.

\subsection{Interfacial deformations }
To parametrize the interface shape $\Sigma_I$ up to orders Ca and $\delta,$ we introduce cylindrical coordinates $(r,\phi,z)$. The relation between Cartesian  and cylindrical coordinates are
\begin{align*}
     x  = r \cos\phi, \quad y = r \sin\phi.
\end{align*}  
The $\mathcal{O}(\mbox{Ca})$ interface shape $h^{(1,0)}$ accounts for the deformation caused by the background flow and satisfies the stress balance equation 
\begin{align}
\nabla^2 h^{(1,0)} -\mbox{Bo} h^{(1,0)} = - \hat{\mathbf{e}}_z\cdot [\bm{\sigma}^{(0,0)} ]\cdot \hat{\mathbf{e}}_z, \label{eqn:single-deformationEqn_Ca}
\end{align}
and the boundary conditions 
\begin{align}
& 0 = - r \frac{\partial h^{(1,0)}}{\partial r} + h^{(1,0)}\quad  \mbox{ at } r  = 1, \label{eqn:single-deformationBC_Ca_r=1}\\
&  h^{(1,0)}\rightarrow 0 \quad \mbox{ as } r \rightarrow \infty. \label{eqn:single-deformationBC_Ca_inf}
\end{align}
The $\mathcal{O}(\delta)$ interface shape $h^{(0,1)}$ accounts for the deformation in the absence of the flow, and it satisfies the stress balance equation 
\begin{align}
\nabla^2 h^{(0,1)} -\mbox{Bo} h^{(0,1)} = 0,\label{eqn:single-deformationEqn_delta}
\end{align}
and the boundary conditions 
\begin{align}
& -\tilde{\theta_s} +\tilde{b} = - r \frac{\partial h^{(0,1)}}{\partial r} + h^{(0,1)}\quad  \mbox{ at } r  = 1, \label{eqn:single-deformationBC_del_r=1} \\
& h^{(0,1)}\rightarrow 0 \quad \mbox{ as } r \rightarrow \infty.  \label{eqn:single-deformationBC_del_inf}
\end{align}

The RHS of Eq. \eqref{eqn:single-deformationEqn_Ca} can be computed from the leading order solutions \eqref{eqn:single-leading-sol-u} and \eqref{eqn:single-leading-sol-p}, and  is given by
    \begin{align}
   &     -\hat{\mathbf{e}}_z\cdot [\bm{\sigma}^{(0,0)}]\cdot \hat{\mathbf{e}}_z =   -(\sigma_{zz}^{2(0,0)} -\sigma_{zz}^{1(0,0)} )  = (1 - \lambda ) \frac{3 y}{2(x^2 + y^2)^{5/2}}  =(1-\lambda) \frac{3 }{2 r^4} \sin\phi.
    \end{align}
To solve for $h^{(1,0)}$, we assume the solution form $h^{(1,0)} = R_P(r)\sin\phi$, then Eq. \eqref{eqn:single-deformationEqn_Ca} can be reduced to a non-homogeneous Bessel's equation in $R_P(r)$, i.e., 
\begin{align}
    r^2 R_P'' + r R_P' - (1+\mbox{Bo}r^2 ) R_P = (1-\lambda)\frac{2}{3 r^2 }.
\end{align}
Using the method of variation of parameters  plus the boundary condition \eqref{eqn:single-deformationBC_Ca_inf}, the solution $R_P$ is obtained. Finally the boundary condition \eqref{eqn:single-deformationBC_Ca_r=1} is applied and we find that  
\begin{align}
\begin{split}
    h^{(1,0)}(r,\phi) 
    & = (1-\lambda)\left[C_1 K_1(\sqrt{\mbox{Bo}}r) - \frac{3}{2} K_1(\sqrt{\mbox{Bo}} r) \int_1^r \frac{1}{r^3} I_1(\sqrt{\mbox{Bo} }r)\mbox{ d}r \right. \\ & \left.-  \frac{3}{2} I_1(\sqrt{\mbox{Bo} }r ) \int_r^{\infty} \frac{1}{r^3} K_1(\sqrt{\mbox{Bo} }r)\mbox{ d}r \right]\sin\phi,
\end{split}   
    \label{eqn:deformationSol}
\end{align}
where 
\begin{align*}
    C_1  = - \frac{3 M  I_2 (\sqrt{\mbox{Bo}})}{2 K_2(\sqrt{\mbox{Bo}})},\quad M =\int_1^\infty \frac{1}{r^3} K_1(\sqrt{\mbox{Bo}}r)\mbox{ d}r
\end{align*}
The solution of \eqref{eqn:single-deformationEqn_delta} with boundary conditions \eqref{eqn:single-deformationBC_del_r=1} and \eqref{eqn:single-deformationBC_del_inf} is 
\begin{align}
    h^{(0,1)} =  \frac{(-\tilde{\theta_s}+\tilde{b})K_0 (\sqrt{\mbox{Bo} }r)}{\sqrt{\mbox{Bo}} K_1(\sqrt{\mbox{Bo}}) + K_0(\sqrt{\mbox{Bo}})}= (-\tilde{\theta}_s + \tilde{b})C_0 K_0(\sqrt{\mbox{Bo} }  r ). \label{eqn:single-staticDeformSol}
\end{align}
Here, $K_n$ and $I_n$ are the modified Bessel functions of order $n$. 
The leading order interfacial deformation is the sum $h = \mbox{Ca} h^{(1,0)} + \delta h^{(0,1)}. $

\begin{figure}
    \centering
    \includegraphics[scale=0.65]{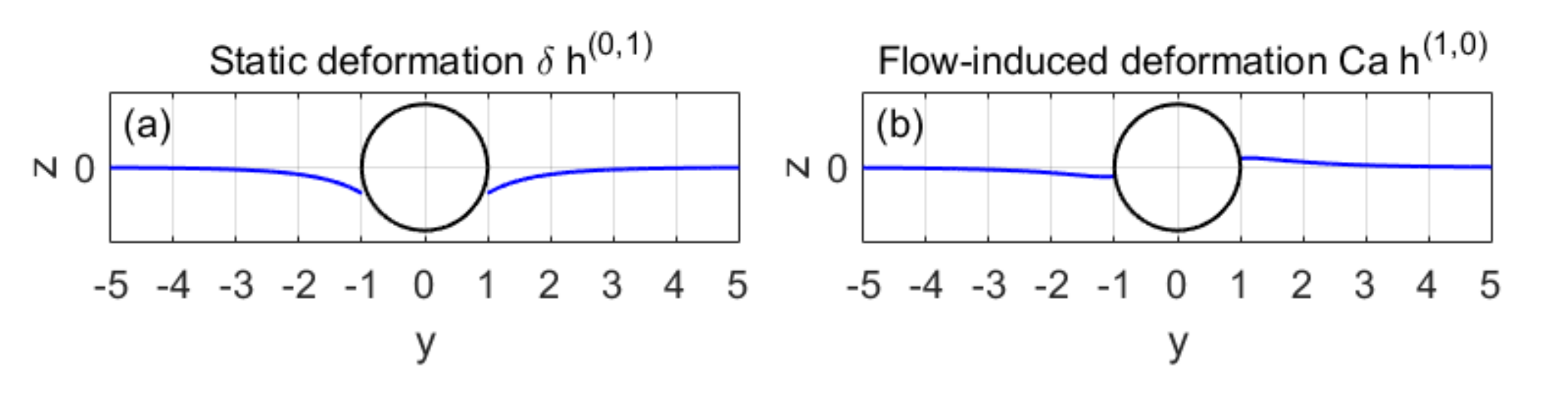}
    \caption{Cross-sections of the static deformation $\delta h^{(0,1)}$ (a) and the flow-induced deformation $\mbox{Ca}h^{(1,0)}$ (b) with $\mbox{Ca} = \delta =\tilde{\theta}_s=1,$ $\mbox{Bo}=1,$ $\lambda=2$, and $\tilde{\theta}_s = 1, \tilde{b}=0.$}
    \label{fig:single-deformations2}
\end{figure}

 Fig. \ref{fig:single-deformations2} shows the $y$-$z$ cross-sections of the static deformation $\delta h^{(0,1)}$ and flow-induced deformation $\mbox{Ca} h^{(1,0)}$ with $\mbox{Ca} = \delta = \tilde{\theta}_s =1,$ $\mbox{Bo}  = 1, \lambda=2$, and $\tilde{b}=0$. The static deformation $\delta h^{(0,1)}$, induced by the contact angle,  describes the equilibrium interface shape in the absence of flow and is axisymmetric; the flow-induced deformation  $\mbox{Ca} h^{(1,0)}$ represents the deformation caused by the uniform background flow $\mathbf{u}^\infty = \hat{\mathbf
e}_y$ and appears anti-symmetric in the $y$ direction. Note that $h^{(0,1)}$ is independent of $\phi$ and $h^{(1,0)}$ depends on $\phi$ as $\sin\phi.$

\begin{figure}
    \centering
    \includegraphics[scale=0.65]{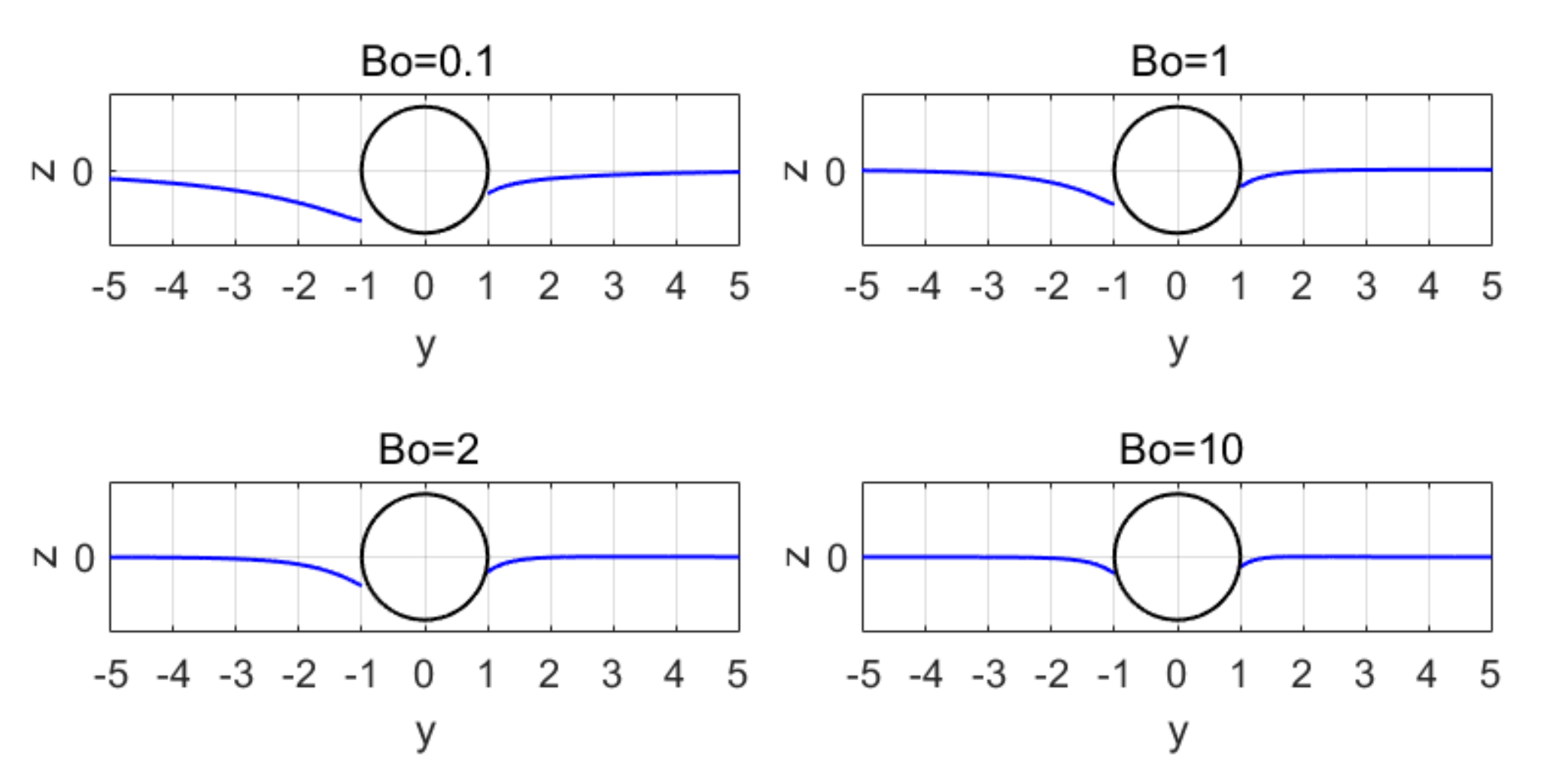}
    \caption{The $y$-$z$ cross-section views of  the interfacial deformation $h = \mbox{Ca} h^{(1,0)} + \delta h^{(0,1)} $ with  $\mbox{Ca} = \delta=\tilde{\theta}_s=1, \lambda = 2$, and varying values of $\mbox{Bo}.$}
    \label{fig:single-deformations1}
\end{figure}

In Fig. \ref{fig:single-deformations1}, we plot the
$y$-$z$ cross-sections of $h = \mbox{Ca} h^{(1,0)} + \delta h^{(0,1)}$ with $\mbox{Ca} = \delta = 1$ and varying values of Bo, from which we see that the amplitude of the deformation decreases as Bo increases, meaning increasing the density mismatch between the two fluid phases flattens the interface shape. Note that in the limit of small Bond number (Bo $\rightarrow 0$), the asymptotic assumption Bo$=\mathcal{O}(1)$ is violated and the deformation solutions become invalid. Also, the values of the parameters chosen in Fig. \ref{fig:single-deformations1} are outside the limits of applicability of our expansions and are chosen to illustrate how the interface is affected by them. Because of this, and since we ignore higher-order terms in the expansions, we observe a mismatch between the fluid interface and the particle surface.

\subsection{Calculation of the drag force}

The drag force exerted on a spherical particle straddling a fluid interface is  
\begin{align}
  F_D = \mathbf{F}_D\cdot (-\mathbf{u}^\infty) = & \sum_{i=1,2} \iint_{\Sigma_{P_i}}  \bm{\sigma}\cdot(- \tilde{\mathbf{n}})\cdot (-\mathbf{u}^\infty)  \mbox{ d}\Sigma. \label{eqn:single-dragFormula}
\end{align}

Inserting the expansions of $\bm{\sigma}$ and $h$ into Eq. \eqref{eqn:single-dragFormula}, we obtain to $\mathcal{O}(\mbox{Ca})$ and $\mathcal{O}(\delta)$
{
\begin{align}
\begin{split}
       F_D 
     = &  \int_{0}^{2 \pi } \left( \int_{0}^{1+\delta b^{(1)}} (\bm{\sigma}^{2(0,0)} + \mbox{Ca} \bm{\sigma}^{2(1,0)} + \delta \bm{\sigma}^{2(0,1)} )\cdot  (\tilde{\mathbf{n} }^{(0,0)} + \delta \tilde{\mathbf{n} }^{(1,0)} )\cdot \mathbf{u}^\infty \mbox{ d}z  \mbox{ d}\phi  \right. \\
     & \left. + \int_{-1+{\delta b^{(1)}}}^0 (\bm{\sigma}^{1(0,0)} + \mbox{Ca} \bm{\sigma}^{1(1,0)} + \delta \bm{\sigma}^{1(0,1)} )\cdot  (\tilde{\mathbf{n} }^{(0,0)} + \delta \tilde{\mathbf{n} }^{(1,0)} )\cdot \mathbf{u}^\infty \mbox{ d}z  \mbox{ d}\phi \right) \\
   & -  \int_{0}^{2 \pi }   \left( \mbox{Ca}  h^{(1,0)}  + \delta  h^{(0,1)}\right)  [\bm{\sigma}^{(0,0)}]\cdot  (\tilde{\mathbf{n} }^{(0,0)} + \delta \tilde{\mathbf{n} }^{(1,0)} )\cdot \mathbf{u}^\infty  \mbox{ d}\phi.
   \end{split} \label{eqn:single-dragFormula-undulation}
     \end{align} } 
To evaluate the surface integrals in Eq. \eqref{eqn:single-dragFormula-undulation}, we introduce spherical coordinates $(\rho,\vartheta,\varphi)$, defined by 
 \begin{align}
x = \rho \sin\varphi \sin\vartheta,\quad       y = \rho \cos\vartheta, \quad z  =  \rho \cos\varphi \sin\vartheta,
\end{align}
where $0 \leq \vartheta \leq \pi$ and $0 \leq \varphi<   2\pi.$

The spherical surface can be described as 
\begin{align}
\Sigma_P = \{(\rho,\varphi,\vartheta) \vert \rho = 1 + \delta \phi^{(1)}(\varphi,\vartheta)\},
\end{align}
where $\phi^{(1)} = \tilde{b}\sin\vartheta\cos\varphi$,
and the unit normal vector $\tilde{\mathbf{n}}$ to the particle surface is
$   \Tilde{\mathbf{n}} =    \Tilde{\mathbf{n}}^{(0,0)} + \delta    \Tilde{\mathbf{n}}^{(0,1)} $ with
\begin{align}
    \Tilde{\mathbf{n}}^{(0,0)} = \hat{\mathbf{e}}_\rho,\quad  \Tilde{\mathbf{n}}^{(0,1)} =-\tilde{b} \cos\vartheta\cos\varphi /\rho  \hat{\mathbf{e}}_\vartheta  +  \tilde{b} \sin\varphi /\rho \hat{\mathbf{e}}_\varphi . 
    \end{align}
    
Substituting in the leading order solutions for flow past a sphere given by Eqs. \eqref{eqn:single-leading-sol-u} and \eqref{eqn:single-leading-sol-p}, we find that 
the $\mathcal{O}(1)$ drag force is the classical result \cite{Maldarelli2022}: 
\begin{align}
    F_D^{(0,0)} = 3 \pi (\lambda + 1).
\end{align}
The formula of the $\mathcal{O}(\mbox{Ca} )$ and $\mathcal{O}(\delta)$ drag forces are given by
\begin{align}
    F_D^{(j,k)} = \underbrace{ \sum_{i=1,2} \iint_{\Sigma_{P_i}^{(0)}}  \bm{\sigma}^{(j,k)}  \cdot \tilde{\mathbf{n}}^{(0,0)}\cdot \mathbf{u}^\infty \mbox{ d}\Sigma}_{ \circled{1}} \underbrace{ - \int_0^{2\pi} h^{(j,k)}  [ \bm{\sigma}^{(0,0)} ] \cdot \tilde{\mathbf{n}}^{(0,0)}\cdot \mathbf{u}^\infty \mbox{ d}\phi}_{ \circled{2}}, \label{eqn:single-correctionDragFormula}
\end{align}
with $(j,k) = (1,0)$ and $(0,1)$,
where $F_D^{(1,0)}$ is the flow-induced correction drag and $F_D^{(0,1)}$ is the contact angle induced correction drag.   
The integral \textcircled{\small 2} in Eq. \eqref{eqn:single-correctionDragFormula} can be evaluated directly with \textcircled{\small 2} $= 0$ for $(j,k) = (1,0)$ and \textcircled{\small 2}$ =-   3 \pi (\lambda-1) (-\tilde{\theta}_s + \tilde{b})C_0 K_0(\sqrt{\mbox{Bo}})$ for $(j,k) = (0,1)$, where $C_0$ is defined in Eq \eqref{eqn:single-staticDeformSol}. The integral in \textcircled{\small 1} containing the correction stress still needs to be evaluated.  

\subsubsection{Lorentz reciprocal theorem}

We use the Lorentz reciprocal theorem to evaluate the term \textcircled{\small 1} in the $\mathcal{O}(\mbox{Ca})$ and $\mathcal{O}(\delta)$ drag formula  \eqref{eqn:single-correctionDragFormula}. 
\begin{figure}
\centering
    \includegraphics[scale=0.45]{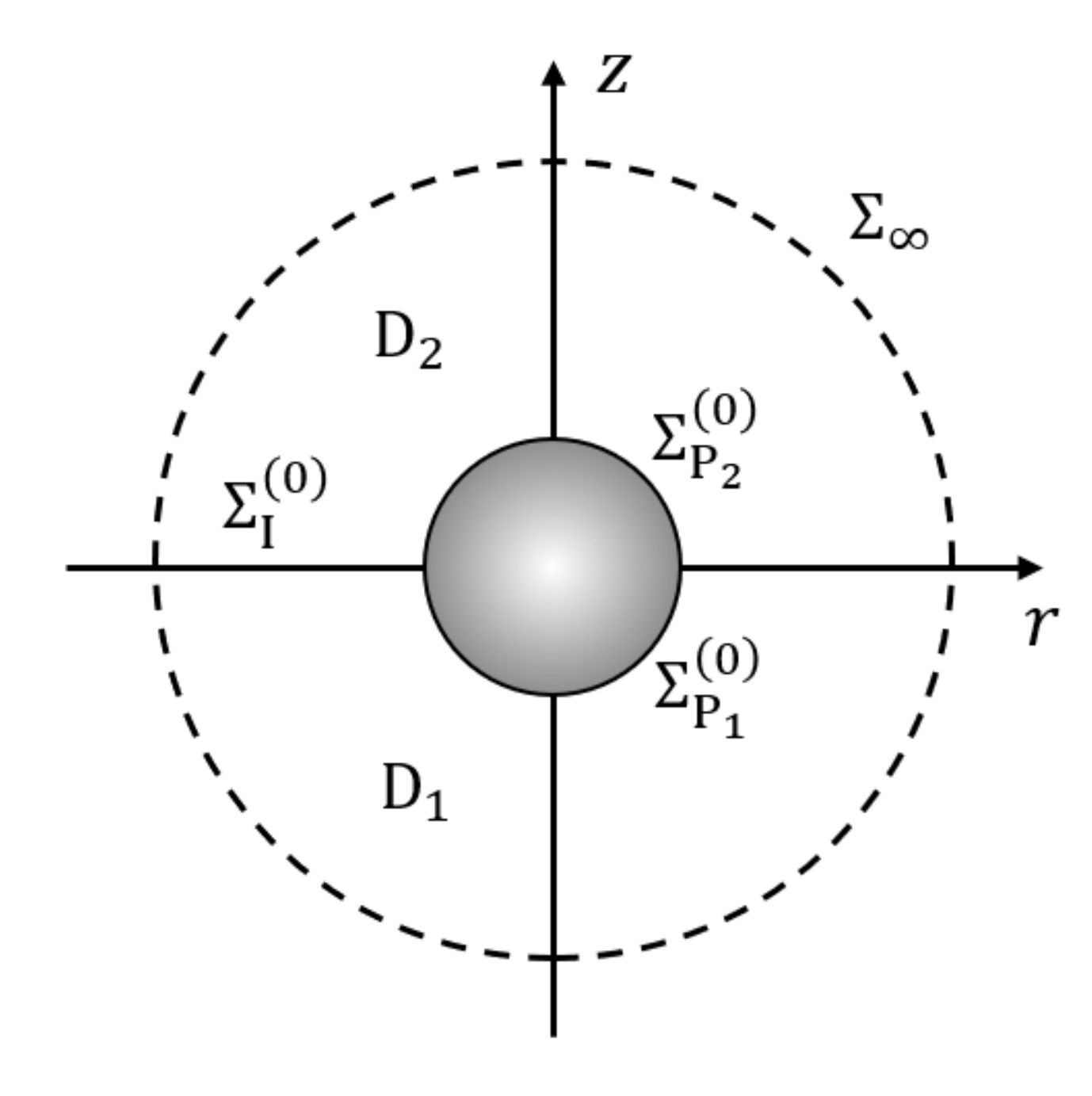}
    \caption{Geometry for the Lorentz reciprocal theorem.}
    \label{fig:single-reciprocal}
\end{figure}
Let $D_i$ denote the region bounded by the particle surface $\Sigma_{P_i}^{(0)}$, the flat interface $\Sigma_I^{(0)}$, and the hemi-spherical surface $\Sigma_{\infty_i }$ at infinity (see Fig. \ref{fig:single-reciprocal}). Let us define $\Sigma_{\infty}  = \Sigma_{\infty_1}  \cup \Sigma_{\infty_2}$.  We apply the Lorentz reciprocal theorem to the following two flow problems: 
\\
\\
\noindent
\textbf{Problem 1:} 
The first problem is constructed by setting Ca and $\delta$ to zero, which describes a sphere bisected by a flat interface in a uniform flow. The flow field and the stress tensor are denoted by $\mathbf{u}^{(0,0)}$ and $\bm{\sigma}^{(0,0)}$, respectively.  We let $\mathbf{u}_D^{(0,0)} = \mathbf{u}^{(0,0)} - \mathbf{u}^\infty$ denote the leading order disturbance field. Then, at the boundaries, 
\begin{align}
&\mathbf{u}_D^{(0,0)} = -\mathbf{u}^\infty \quad \mbox{ for } \mathbf{x} \in \Sigma_{P_i}^{(0)}, \\
&\mathbf{u}_D^{(0,0)} = \mathbf{0} \quad \mbox{ for }\mathbf{x} \in \Sigma_\infty.
\end{align} 
\\
\\
\noindent
\textbf{Problem 2:} The second problem is described by the truncated asymptotic expansions in $\mbox{Ca}$ and $\delta$:
\begin{align}
\mathbf{u}_D = & \mathbf{u}_D^{(0,0)}+ \mbox{Ca}  \mathbf{u}^{(1,0)}_D + \delta \mathbf{u}^{(0,1)}_D,\\
\bm{\sigma} = & \bm{\sigma}^{(0,0)} + \mbox{Ca}   \bm{\sigma}^{(1,0)} +  \delta \bm{\sigma}^{(0,1)},
\end{align}
where $\mathbf{u}_D$ denotes the disturbance field.
For $\mathbf{x} \in \Sigma_\infty $,
\begin{align}
\mathbf{u}_D = 0,
\end{align}
for $\mathbf{x} \in \Sigma^{(0)}_{P_i},$
\begin{align}
& \mathbf{u}^{(1,0)}_D = -\mathbf{u}^\infty,\quad    \mathbf{u}^{(1,0)}_D = \mathbf{0},
  \quad  \mathbf{u}^{(0,1)}_D = -\tilde{b}\frac{\partial \mathbf{u}^{(0,0)}_D}{\partial z },\label{eqn:single-reciprocal-bc1}
  \end{align}
and for $\mathbf{x} \in \Sigma^{(0)}_I$,
\begin{align}
    & \mathbf{u}_D^{(0,0)} \cdot \hat{\mathbf{e}}_z = 0, \label{eqn:single-reciprocal-ic1}\\
    & \mathbf{u}_D^{(1,0)} \cdot \hat{\mathbf{e}}_z = -\mathbf{u}_D^{(0,0)}\cdot \hat{\mathbf{n}}^{(1)} - \frac{\partial \mathbf{u}_D^{(0,0)}}{\partial z} h^{(1,0)}\cdot \hat{\mathbf{e}}_z ,\label{eqn:single-reciprocal-ic2}\\ 
  &    \hat{\mathbf{t}}^{(0,0)} \cdot  [\bm{\sigma}^{(1,0)}] \cdot \hat{\mathbf{e}}_z =  - \hat{\mathbf{t}}^{(0)} \cdot  \frac{\partial [\bm{\sigma}^{(0,0)}]}{\partial z }h^{(1,0)} \cdot \hat{\mathbf{e}}_z - \hat{\mathbf{t}}^{(0,0)} \cdot [\bm{\sigma}^{(0,0)}] \cdot \hat{\mathbf{n}}^{(1,0)} - \hat{\mathbf{t}}^{(1,0)} \cdot [\bm{\sigma}^{(0,0)}] \cdot \hat{\mathbf{e}}_z,\label{eqn:single-reciprocal-ic3}
\end{align}
and similarly for $\mathbf{u}^{(0,1)}\cdot \hat{\mathbf{e}}_z$ and $ \hat{\mathbf{t}}^{(0,0)} \cdot  [\bm{\sigma}^{(0,1)}] \cdot \hat{\mathbf{e}}_z$.

Since solutions $(\mathbf{u}_D^{(0,0)}, \bm{\sigma}^{(0,0)})$ and $(\mathbf{u}_D, \bm{\sigma})$ are defined in the same geometry, they are related by the reciprocal theorem
\begin{align}
\iint_{\Sigma_i}(\bm{\sigma}^{(0,0)} \cdot \mathbf{n}) \cdot \mathbf{u}_D \mbox{ d}\Sigma = \iint_{\Sigma_i} (\bm{\sigma} \cdot \mathbf{n}) \cdot \mathbf{u}^{(0,0)}_{D} \mbox{ d}\Sigma,\quad i = 1,2, \label{eqn:single-reciprocalEqn1}
\end{align}
where $\Sigma_i =\partial D_i =  \Sigma^{(0)}_{P_i} \cup \Sigma^{(0)}_{I_i} \cup \Sigma_{\infty_i}$, and $\mathbf{n}$ denote the outward normal of $\Sigma_i$.
The contribution from the far-field integral vanishes, since
\begin{align}
\vert \vert  \mathbf{u}_D^{(0,0)} \vert \vert \sim \vert \mathbf{x}\vert^{-1}, \quad \vert \vert  \mathbf{u}_D\vert \vert \sim \vert \mathbf{x}\vert^{-1}, \quad  \vert \vert  \bm{\sigma}^{(0,0)}\cdot \mathbf{n} \vert \vert \sim \vert \mathbf{x}\vert^{-2}, \quad \mbox{ and }  \vert \vert  \bm{\sigma}\cdot \mathbf{n} \vert \vert \sim \vert \mathbf{x}\vert^{-2}.
\end{align}

Collecting coefficients of $\mbox{Ca}$ and $\delta$ in Eq. \eqref{eqn:single-reciprocalEqn1}, we are able to express terms \textcircled{\small 1} in Eq. \eqref{eqn:single-correctionDragFormula} in terms of integrals over the flat interface $\Sigma_{I}^{(0)}$, i.e.,
\begin{align}
\begin{split}
\mbox{\textcircled{\small 1}}
 & =  \iint_{\Sigma^{(0)}_I} [\bm{\sigma}^{(0,0)}] \cdot (-\hat{\mathbf{e}}_z) \cdot  \mathbf{u}_D^{(1,0)}  \mbox{ d}\Sigma  -\iint_{\Sigma^{(0,0)}_I} [\bm{\sigma}^{(1,0)}]  \cdot (-\hat{\mathbf{e}}_z) \cdot  \mathbf{u}^{(0,0)}_D  \mbox{ d}\Sigma,
\end{split} \label{eqn:single-reciprocal_drag_10}
\end{align}
for the $\mathcal{O}(\mbox{Ca})$ drag $F_D^{(1,0)}$, and
\begin{align}
\begin{split}
\mbox{ \textcircled{\small 1}} 
 = &   \iint_{\Sigma^{(0)}_I} [\bm{\sigma}^{(0,0)}] \cdot (-\hat{\mathbf{e}}_z) \cdot  \mathbf{u}_D^{(0,1)}  \mbox{ d}\Sigma  -\iint_{\Sigma^{(0,0)}_I} [\bm{\sigma}^{(0,1)}]  \cdot (-\hat{\mathbf{e}}_z) \cdot \mathbf{u}^{(0,0)}_D \mbox{ d}\Sigma \\
  &-\sum_{i=1,2}\iint_{\Sigma^{(0)}_{P_i}} \bm{\sigma}^{(0,0)} \cdot (-\tilde{\mathbf{n}}^{(0,0)} ) \cdot \mathbf{u}^{(0,1)} \mbox{ d}\Sigma
\end{split} \label{eqn:single-reciprocal_drag_01}
\end{align}
for the $\mathcal{O}(\delta)$ drag $F_D^{(0,1)}$, where the correction velocities and stress jumps are given in the correction boundary conditions (see Appendix \ref{app:single-leading-corr-problems})

The additional drag contribution when the particle's center is shifted from the origin is
\begin{align}
    -\sum_{i=1,2}\iint_{\Sigma^{(0)}_{P_i}} \bm{\sigma}^{(0,0)} \cdot (-\tilde{\mathbf{n}}^{(0,0)} ) \cdot \mathbf{u}^{(0,1)} \mbox{ d}\Sigma = \frac{27}{16} \pi (\lambda-1)\tilde{b}.
\end{align}
The result recovers the correction drag in Eq. (3.17) from \cite{dorr2015}, where the particle translates along a flat gas-liquid interface ($\lambda = \mu_2/\mu_1 = 0$) with immersion depth $\delta \tilde{b}.$

\begin{figure}
        \centering
        \includegraphics[scale=0.65]{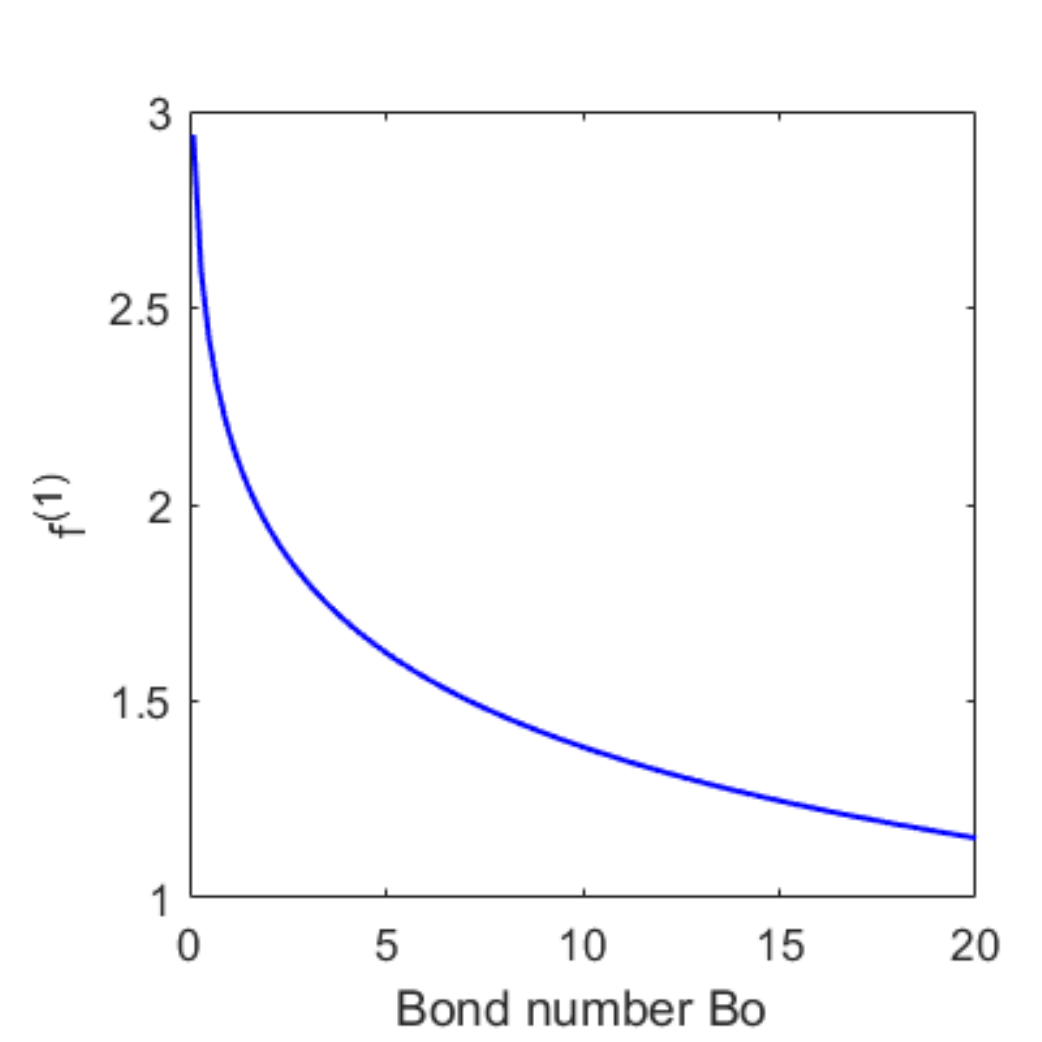}
            \caption{The drag coefficient as a function of Bond number Bo. }
            \label{fig:single-dragCoeff}
\end{figure}

Continuing, we can now calculate the first correction to the drag. Because of the anti-symmetry of the flow field given by Eqs. \eqref{eqn:app-single-leading-ur} - \eqref{eqn:app-single-leading-uz}, the flow induced drag $F^{(1,0)}$ is zero. The contact angle induced correction drag is $F_D^{(0,1)}$ is given by 
    \begin{align}
\begin{split}
 F_D^{(0,1)} =  & \pi \int_1^\infty \left(  -B_z(r)  \frac{3(\lambda-1)}{2r^4} +  B_r(r) (\tilde{u}_r^{(0,0)} (r)-1)  + B_\phi(r) (\tilde{u}_\phi^{(0,0)}(r) -1) \right) r \mbox{ d}r \\
& + \frac{27}{16} \pi(\lambda-1)\tilde{b} - 3\pi (\lambda -1)(-\tilde{\theta}_s + \tilde{b})C_0 K_0(\sqrt{\mbox{Bo}}), 
\end{split}
\label{eqn:single-correctionDragFormula_01-2}
\end{align} 
Here,
\begin{align}
&  \tilde{u}_r^{(0,0)}  = \frac{1}{4}\left(  \frac{-6 r^2 + 2 }{r^3} + 4 \right),\quad  \tilde{u}_\phi^{(0,0)} = \frac{1}{4} \left( -\frac{3}{r} - \frac{1}{r^3} + 4  \right), \quad  \frac{\partial \tilde{u}^{(0,0)}_z }{\partial z} = \frac{3(1-r^2)}{4 r^4},\\
& B_z= (\tilde{u}^{(0,0)}_r-1) \frac{\mbox{d} h^{(0,1)} }{\mbox{d}r}  - \frac{\partial \tilde{u}^{(0,0)}_z }{\partial z} h^{(0,1)},\\
& B_r = \frac{3(\lambda-1)}{2r^5} \left( (4-3r^2)h^{(0,1)} + 3 r(r^2-1)\frac{\mbox{d} h^{(0,1)} }{\mbox{d}r}  \right),\\
& B_\phi = \frac{3(\lambda-1)}{2 r^5} (r \frac{\mbox{d} h^{(0,1)} }{\mbox{d}r}  - h^{(0,1)}).
\end{align}
The contact angle induced drag \eqref{eqn:single-correctionDragFormula_01-2} is numerically evaluated using the trapezoidal rule. 
We can now write the total drag to $\mathcal{O}(\mbox{Ca})$ and $\mathcal{O}(\delta)$ as 
\begin{align}
    F_D = &  3\pi (\lambda+1) + \delta (\lambda-1) \left( (\tilde{\theta}_s - \tilde{b})f^{(1)}(\mbox{Bo}) + \frac{27}{16} \pi \tilde{b}\right)  \label{eqn:single-truncatedDrag}
\end{align}
where $f^{(1)}$, shown in Fig. \ref{fig:single-dragCoeff}, is the correction drag coefficient in terms of Bo. Recall that increasing Bo, which represents the density mismatch between the two fluid phases, flattens the interface shape near the particle. Consequently, an increase in Bo (e.g., by increasing the density mismatch) reduces the correction drag force caused by interfacial deformation, as shown in Fig. \ref{fig:single-dragCoeff}. 
In Eq. \eqref{eqn:single-truncatedDrag}, we observe that the correction drag force at order $\delta$ scales linearly with the viscosity difference $(\lambda-1)$, and when the two fluid phases have the same viscosity ($\lambda$=1), the $\mathcal{O}(\delta)$ drag vanishes. This vanishing of the correction drag when $\lambda=1$ is not expected if higher order terms were included.

\begin{figure}
\centering
\includegraphics[scale=0.65]{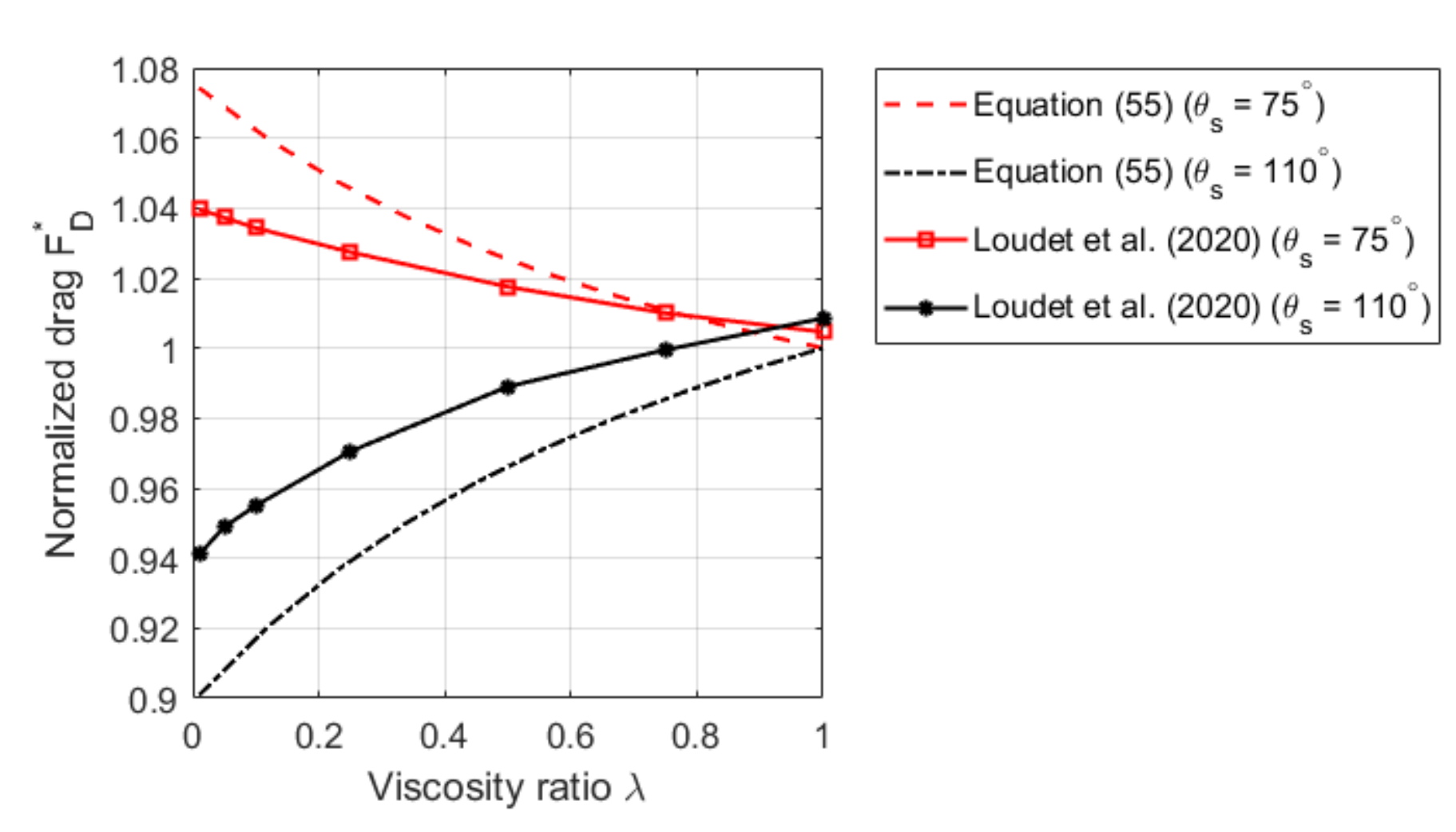}
 \caption{Comparison of the normalized drag $F_D^*$ with numerical results from Loudet et al. (2020), for contact angles $\theta_s = 75^\circ$ and $110^\circ,   \mbox{Bo}\approx 0.2, \tilde{b} = 0. $ }
\label{fig:single-loudet-fig7}
\end{figure}

\begin{figure}
\centering
\includegraphics[scale=0.65]{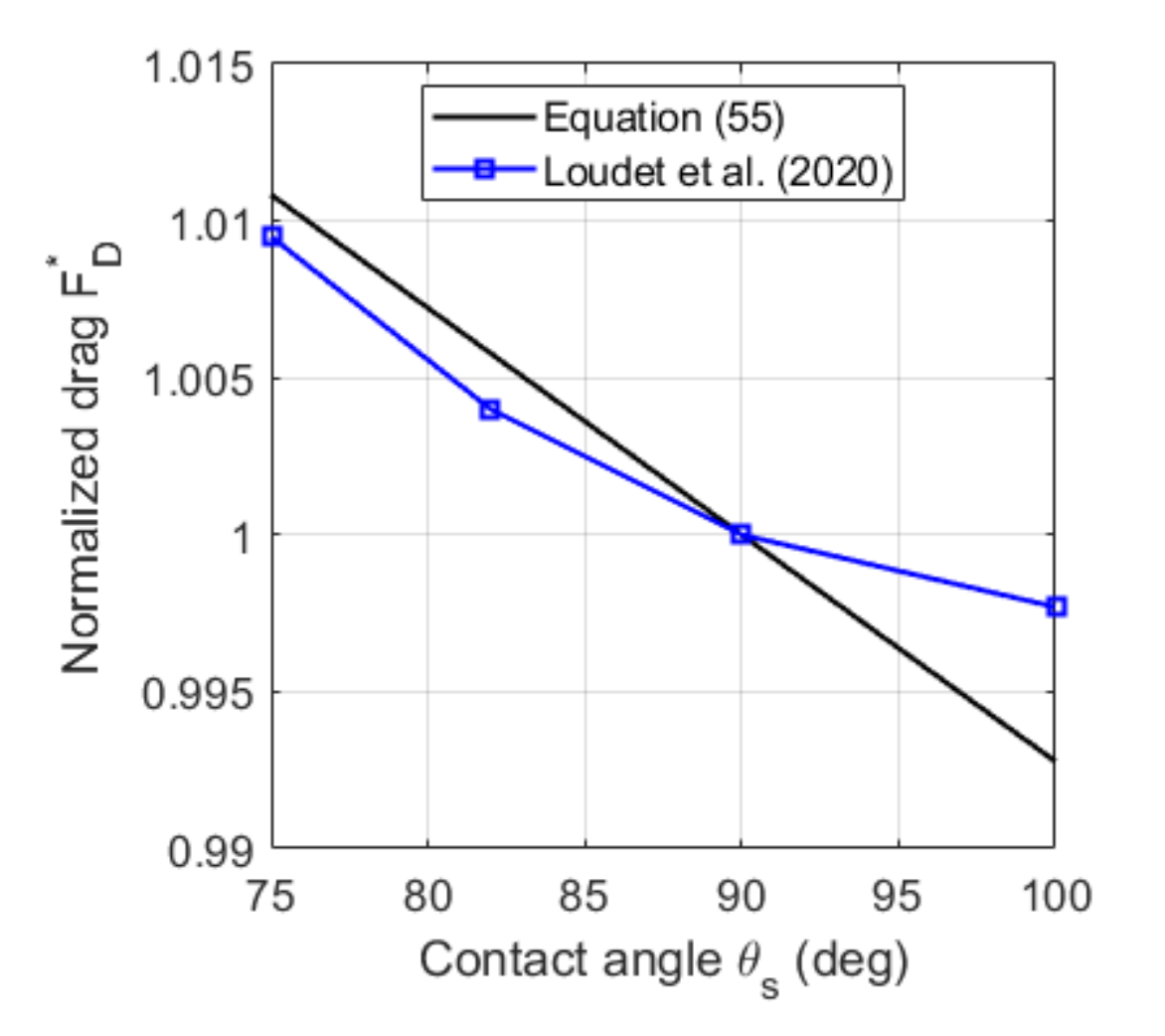}
 \caption{Comparison of the normalized drag $F_D^*$  with numerical results from Loudet et al. (2020) for viscosity ratio $\lambda = 0.75, \mbox{Bo} \approx 0.2,$ and $\tilde{b}=0$. }
\label{fig:single-loudet-fig6}
\end{figure}

In Figs.  \ref{fig:single-loudet-fig7} and \ref{fig:single-loudet-fig6}, we compare the normalized drag
\begin{align}
    F_D^* = \frac{F_D}{3\pi (\lambda+1)} \label{eqn:single-normalizedDrag} 
\end{align}
with the 2D numerical results of Loudet et al. \cite{Loudet2020}. The drag forces in \cite{Loudet2020} are calculated with $\mbox{Ca} \sim 10^{-3}-10^{-4}$ and $40^\circ<\theta_s <140^\circ$.
In Fig. \ref{fig:single-loudet-fig7}, we plot the normalized drag as a function of viscosity ratio $\lambda$ in comparison with Loudet et al.'s results for contact angle $\theta_s = 75^\circ$ and $\theta_s = 110^\circ.$ We see that the asymptotic solutions are qualitatively consistent with the numerical results in that as the viscosity ratio $\lambda$ tends to 1, the effect of deformations on the drag force decreases. Quantitative differences are observed. There could be several reasons for this. First, the comparisons are made between a 3D flow in an unbounded Stokes fluid and a 2D Navier-Stokes flow confined between two parallel planes. Second, the values of the contact angles, $\theta_s = 75^\circ$ and $\theta_s = 110^\circ$, violates the assumption of small correction contact angles for the asymptotic expansion. In Fig. \ref{fig:single-loudet-fig6}, we set $\lambda = 0.75$ and compare the predictions. The asymptotic solution predicts the drag dependence on $\theta_s$ to first order in $\delta$ (linear effects), and \cite{Loudet2020}'s solution to the full flow problem captures the contact angle's higher-order nonlinear effects on the drag force. 
    
\section{Pair interactions of particles}
In this section, we consider the steady motion of two spherical particles at a fluid interface under creeping flow conditions, where the background flow is arbitrarily oriented relative to the spheres' line-of-centers. The linearity of the Stokes equations and the boundary conditions allows us to decompose the problem into two sub-problems: uniform flows past two spheres at an interface, where the imposed flow direction is either perpendicular or parallel to the spheres' line-of-centers. 
 
To move forward, we employ the solutions for the motion of two spheres in an unbounded fluid \cite{Stimson1926,Goldman1966}. Stimson and Jeffery \cite{Stimson1926} solved the problem of two spheres translating with a constant velocity parallel to their line-of-centers. Goldman et al. \cite{Goldman1966} calculated the terminal setting motion of two arbitrarily oriented spheres by combining Stimson and Jeffery's solutions \cite{Stimson1926} with the solutions to the side-by-side problem, in which the motion of the spheres is perpendicular to their line-of-centers. 

Using the same approach as for the single-particle problem, we study the influence of interfacial deformations on the drag force acting on the particles, where the solutions obtained by Goldman et al. \cite{Goldman1966} and Stimson and Jeffery \cite{Stimson1926} are used to solve the leading order problems. 

\subsection{Flow perpendicular to the particles' line-of-centers }
\subsubsection{Problem formulation}
\begin{figure}
        \centering
        \includegraphics[width=11cm]{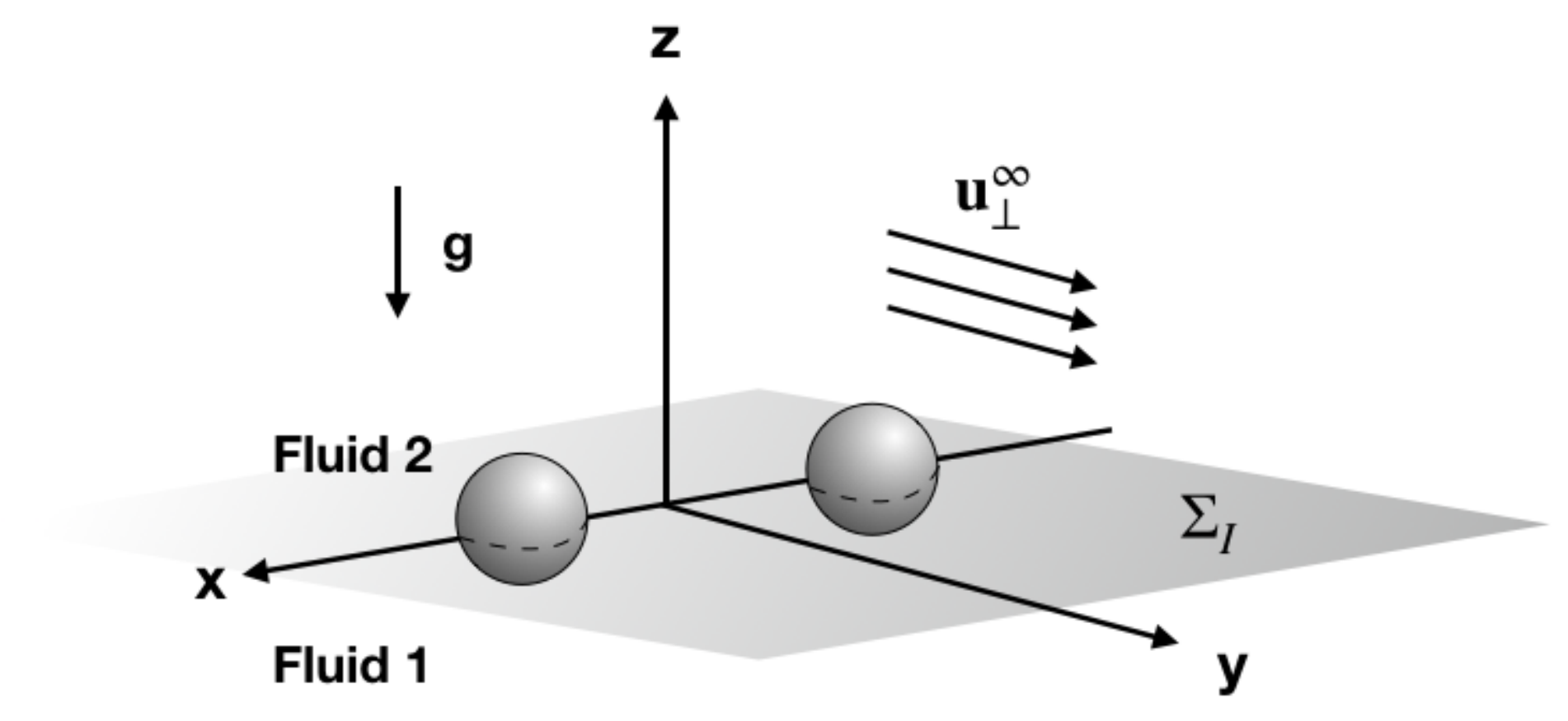}
        \caption{Illustration of two spherical particles straddling a fluid interface between two immersible viscous fluids, where the background uniform flow is perpendicular to the line-of-centers of the two particles.}
        \label{fig:perp-setup1}
    \end{figure}

Consider two spherical particles of radii $a$ straddling a fluid interface between two viscous fluids with respective viscosities $\mu_1$ and $\mu_2$ in a uniform flow perpendicular to the line-of-centers of the two spheres (see Fig. \ref{fig:perp-setup1}). We assume the two particles have their centers of masses pinned at $-L/2\hat{\mathbf{e}}_x$ and $L/2 \hat{\mathbf{e}}_x$, respectively, where $L$ denotes the dimensionless separation distance between the two particles. The nondimensionalized background flow is denoted by $\mathbf{u}^\infty_\perp = \hat{\mathbf{e}}_y,$ and we adopt similar notations used in the previous section.   
    
As in section \ref{sect:single-particle}, we perturb the contact angle $\theta_s$ from $90^\circ$, i.e., $\theta_s = \pi/2 + \delta\tilde{\theta}_s$. 
 For any nondimensionalized quantity $f_\perp$, where the subscript $``\perp"$ indicates variables in the perpendicular flow problem, we consider the two-parameter asymptotic expansion for Ca and $\delta$: 
\begin{align}
    f_\perp = f_\perp^{(0,0)} + \mbox{Ca} f_\perp^{(1,0)} + \delta f_\perp^{(0,1)} + \cdots. \label{eqn:perp-asym-exp}
\end{align}
In this discussion, we do not perturb the height of the center of the sphere. This could be easily included by paralleling the analysis of section \ref{sect:single-particle}.  

The leading order problem describes two spheres at a flat fluid interface in a perpendicular uniform flow.  The equivalent problem of two spheres translating at a constant velocity perpendicular to their line-of-centers in a viscous fluid in the absence of an interface is solved by Goldman et al. \cite{Goldman1966}. The analytical solutions of the pressure and velocity field in \cite{Goldman1966} are the leading order pressure $p^{(0,0)}_\perp$ and velocity field $\mathbf{u}^{(0,0)}_\perp$ (see Appendix \ref{app:perp-leading-sol}). As in section \ref{sect:single-particle}, we modified these solutions to account for the different viscosities of the two fluids. 

To better describe the two-sphere geometry, we introduce bicylindrical coordinates $(\sigma,\tau,z)$. The relations between the Cartesian coordinates and the bicylindrical coordinates are 
\begin{align}
    x = \frac{c\sinh\tau}{\cosh\tau -\cos\sigma}, \quad y = \frac{c\sin\sigma}{\cosh\tau - \cos\sigma},\quad -\pi < \sigma <\pi, \quad -\tau_1 < \tau < \tau_1,
    \label{eqn:perp-bicylindricalCoordinates}
\end{align}
where $c = \sqrt{(L/2)^2-1}$ is the separation coefficient, and $\tau = \pm \tau_1 = \pm \mbox{arccosh}(L/2)$ describe the TCLs at the particle surfaces at leading order. 

\subsubsection{Interfacial deformations}
Paralleling the analysis of section \ref{sect:single-particle}, the $\mathcal{O}(\mbox{Ca})$ deformation $h_\perp^{(1,0)}$ and $\mathcal{O}(\delta)$ deformation $h^{(0,1)}_\perp$ satisfy the stress balance equations
\begin{align}
    \nabla^2 h^{(1,0)}_\perp -\mbox{Bo} h^{(1,0)}_\perp = -\hat{\mathbf{e}}_z \cdot [\bm{\sigma}_\perp^{(0,0)}]\cdot \hat{\mathbf{e}}_z, \label{eqn:perp-stressBalanceEq_10} 
\end{align}
    and 
\begin{align}
    \nabla^2 h^{(0,1)}_\perp -\mbox{Bo} h^{(0,1)}_\perp = 0.\label{eqn:perp-stressBalanceEq_01} 
\end{align}
The normal-normal stress difference $ \hat{\mathbf{e}}_z \cdot [\bm{\sigma}_\perp^{(0,0)}]\cdot \hat{\mathbf{e}}_z$ can be calculated from the leading order solutions $p^{(0,0)}_\perp$ and $\mathbf{u}_\perp^{(0,0)}$. Then, Eq. \eqref{eqn:perp-stressBalanceEq_10} in bicylindrical coordinates reads
\begin{align}
    \frac{(\cosh\tau - \cos\sigma)^2 }{c^2 } \left( \frac{\partial^2 h_\perp^{(1,0)}}{\partial \sigma^2 } + \frac{\partial^2 h_\perp^{(1,0)}}{\partial \tau^2 }\right) -\mbox{Bo} h_\perp^{(1,0)} = -\frac{2(\lambda-1)X(\sigma,\tau)}{c \sin\sigma/(\cosh\tau - \cos\sigma)}, \label{eqn:perp-stressBalanceEq_10_bicyl}
\end{align}
with
\begin{align}
     X = &  (\cosh\tau - \cos\sigma)^{1/2} \sin^2\sigma \sum_{n=2}^\infty  F_n \cosh(n+1/2)\tau P_n''(\cos\sigma),
\end{align}
where $P_n$ denotes the Legendre polynomial of order $n$, and the coefficients $F_n$ are given in Eqs. (3.55) and (3.56) in \cite{Goldman1966} and included in Appendix \ref{app:perp-leading-sol} (see \cite{zhou2022} for further details).
Likewise, the stress balance equation \eqref{eqn:perp-stressBalanceEq_01} is given by 
\begin{align}
    \frac{(\cosh\tau - \cos\sigma)^2 }{c^2 } \left( \frac{\partial^2 h_\perp^{(0,1)}}{\partial \sigma^2 } + \frac{\partial^2 h_\perp^{(0,1)}}{\partial \tau^2 }\right) -\mbox{Bo} h_\perp^{(0,1)} = 0.
\end{align}

\begin{figure}
        \centering
        \includegraphics[scale=0.4]{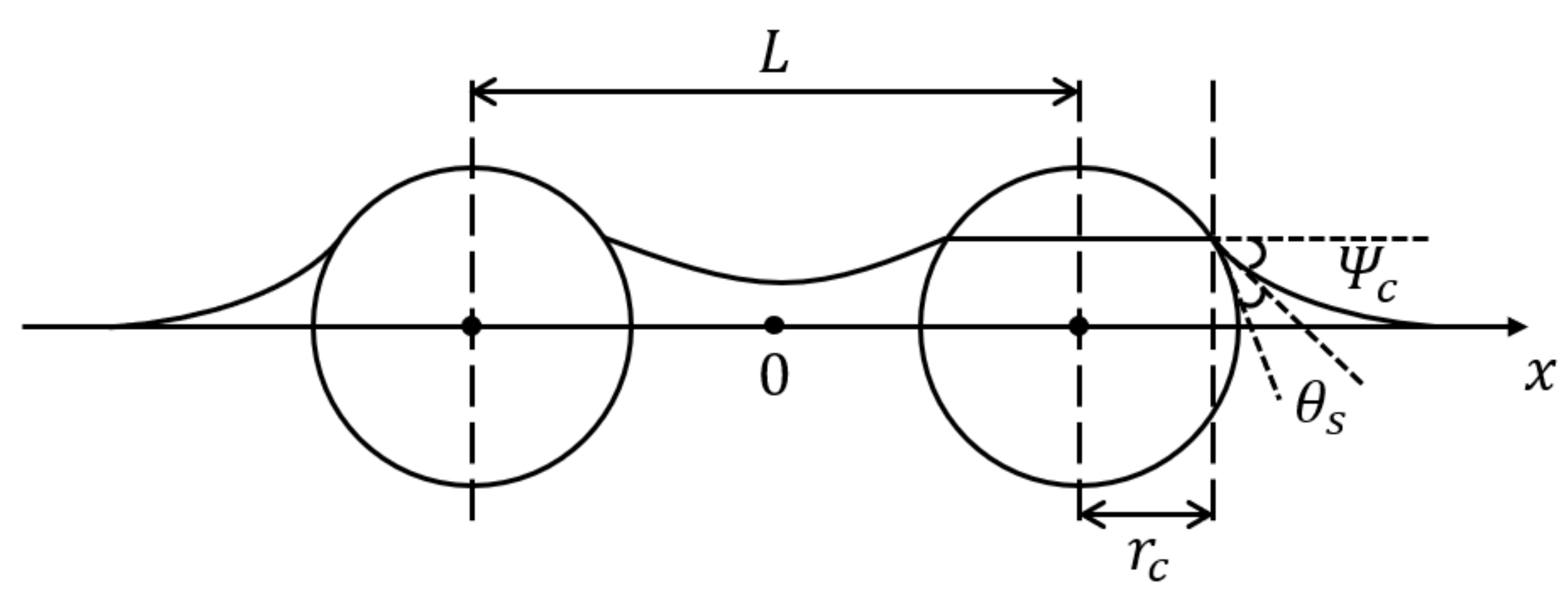}
        \caption{Sketch of the fluid interface near two spherical particles.}
        \label{fig:perp-contact-angle}
    \end{figure}

\begin{figure}
    \centering
    \includegraphics[scale=0.68]{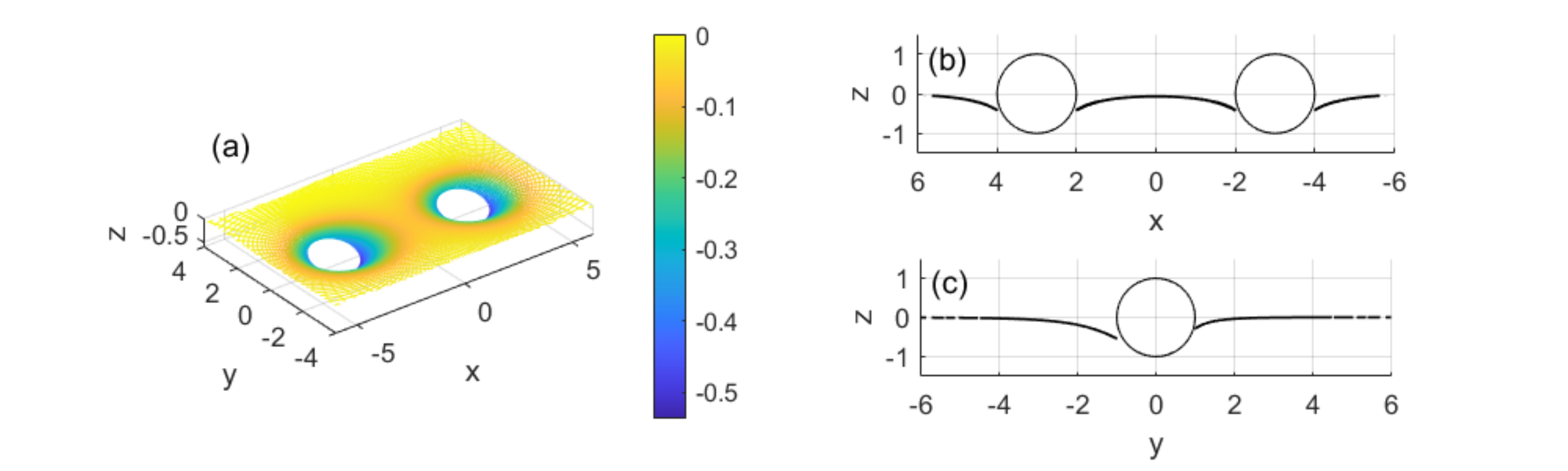}
    \caption{The numerically calculated interfacial deformation, $h = \mbox{Ca}h_\perp^{(1,0)} +\delta h_\perp^{(0,1)}$, around two spherical particles with $\mbox{Ca} = \delta=\tilde{\theta}_s =1$, Bo $=1, \lambda=2$ and $L = 6$: (a) 3D visualization, where the colormap shows the interfacial height near the particle (b) $x$-$z$ cross-section view for $y = 0$, (c) $y$-$z$ cross-section view for $x = L/2$.   }
    \label{fig:perp-deformations1-2}
\end{figure}

The unperturbed fixed contact angle conditions at the TCL are 
\begin{align}
    \cos(\pi/2 - \Psi_c) = \pm \hat{\mathbf{e}}_\tau \cdot \hat{\mathbf{n}}_\perp \big\vert_{\tau =\pm \tau_1},\label{eqn:perp-contact-angle1}
\end{align}
where $\Psi_c$ is the inclination angle, $\hat{\mathbf{e}}_\tau$ is the unit tangent to the $\tau$ contour lines, and $\hat{\mathbf{n}}_\perp \big\vert_{\tau =\pm \tau_1}$ is the unit normal to the fluid interface evaluated at the TCLs (see Fig. \ref{fig:perp-contact-angle}). Substituting the asymptotic expansions into Eq. \eqref{eqn:perp-contact-angle1} and expanding in Ca and $\delta$ yields the boundary conditions for $h_\perp^{(1,0)}$ and  $h_\perp^{(0,1)}$ 
\begin{align}
 &  \left.  \left[   \pm \frac{\cosh\tau-\cos\sigma}{c}\frac{\partial h_\perp^{(1,0)}}{\partial \tau} + h_\perp^{(1,0)} \right] \right\vert_{\tau = \pm\tau_1} = 0,\label{eqn:perp-h10-bc}\\
 &  \left.  \left[   \pm \frac{\cosh\tau-\cos\sigma}{c}\frac{\partial h_\perp^{(0,1)}}{\partial \tau} + h_\perp^{(0,1)} \right] \right\vert_{\tau = \pm\tau_1} = -\tilde{\theta}_s \label{eqn:perp-h01-bc},
\end{align}
respectively (see Appendix \ref{app:perp-contact-angle-cond} for details). It should be noted that the PDEs \eqref{eqn:perp-stressBalanceEq_10} and \eqref{eqn:perp-stressBalanceEq_01} are defined on the rectangular region $|\tau| < \tau_1$ and $|\sigma|<\pi.$ These second order linear PDEs can be solved with a straightforward centered finite difference scheme, along with second order finite difference approximations of Eqs. \eqref{eqn:perp-h10-bc} and \eqref{eqn:perp-h01-bc}, plus the periodic conditions at $\sigma = \pm \pi.$ The resulting linear system is solved using MATLAB's backslash operator. Second order convergence is observed. The partial derivatives of $h_\perp^{(1,0)}$ and $h_\perp^{(0,1)}$  with respect to $x$ and $y$ are obtained using finite difference approximations, which have linear convergence. Fig. \ref{fig:perp-deformations1-2} shows the numerically evaluated interfacial deformation $h = \mbox{Ca}h_\perp^{(1,0)} +\delta h_\perp^{(0,1)}$ around two spherical particles and its cross-section plots with $\mbox{Ca} =1, \delta = \tilde{\theta}_s =1$, Bo $=1,$ and $L = 6.$ As in Figs. \ref{fig:single-deformations2} and \ref{fig:single-deformations1}, we set $\mbox{Ca} = \delta = \theta_s = 1$ to illustrate the deformations. 

\subsubsection{Calculation of the drag force}

We let $\Sigma_{P_{i}^\text{\tiny I}}$ and $\Sigma_{P_{i}^\text{\tiny II}}$ denote the respective surfaces of the two particles in fluid phase $i$.
Then, the total drag force acting on the two particles is defined as
\begin{align}
    F_\perp = \mathbf{F}_\perp \cdot (-\mathbf{u}^\infty_\perp) = & \sum_{i=1,2} \iint_{\Sigma_{P^\text{\tiny I}_i}}  \bm{\sigma}_\perp\cdot(- \tilde{\mathbf{n}}_\perp)\cdot (-\mathbf{u}_\perp^\infty)  \mbox{ d}\Sigma +  \iint_{\Sigma_{P^\text{\tiny II}_i}}  \bm{\sigma}_\perp\cdot(- \tilde{\mathbf{n}}_\perp)\cdot (-\mathbf{u}_\perp^\infty)  \mbox{ d}\Sigma. \label{eqn:perp-dragFormula1}
\end{align}
By symmetry arguments, the drag forces on the two spheres are equal and the drag on each sphere is $F_\perp/2.$ Paralleling the drag force calculation \eqref{eqn:single-dragFormula-undulation} for a single sphere, we find that to order Ca and $\delta$,  $F_\perp = F_\perp^{(0,0)} + \mbox{Ca} F_\perp^{(1,0)} + \delta F_{\perp}^{(0,1)},$
where the $\mathcal{O}(1)$ drag force
\begin{align}
   F_\perp^{(0,0)}  =  \sum_{i=1,2} \iint_{\Sigma^{(0)}_{P_i^{\text{\tiny I,II}}}}   \bm{\sigma}_\perp^{(0,0)} \cdot \tilde{\mathbf{n} }_{\perp} ^{(0,0)} \cdot \mathbf{u}_\perp^\infty \mbox{ d}\Sigma 
\end{align}
is computed by Goldman et al. \cite{Goldman1966}. The $\mathcal{O}(\mbox{Ca})$ and $\mathcal{O}(\delta)$ drag forces, $F_\perp^{(1,0)}$ and $F_\perp^{(0,1)}$, are defined as
\begin{align}
   &  F_\perp^{(j,k)}  = \sum_{i=1,2} \iint_{\Sigma^{(0)}_{P_i^{\text{\tiny I,II}}}}   \bm{\sigma}_\perp^{(j,k)} \cdot \tilde{\mathbf{n} }_{\perp} ^{(0,0)} \cdot \mathbf{u}_\perp^\infty \mbox{ d}\Sigma -  \int_{\Sigma^{(0)}_{\text{\scriptsize TLC}^{\text{\tiny I,II}}}}    h_\perp^{(j,k)}  [\bm{\sigma}_\perp^{(0,0)}]\cdot \tilde{\mathbf{n} }_{\perp} ^{(0,0)}\cdot \mathbf{u}_\perp^\infty  \mbox{ d}s,\label{eqn:perp-corrDrag}
\end{align}
with $(j,k) = (1,0)$ and $(0,1)$, where the integrals can be evaluated over particle I and II with 
\begin{align*}
    \Sigma^{(0)}_{P_i^{\text{\tiny I,II}}} = \{(x,y,z)\vert (x\pm L/2)^2 + y^2 +z^2=1 \}, \quad \mbox{ and }\Sigma^{(0)}_{\text{\scriptsize TLC}^{\text{\tiny I,II}}} =\{(\sigma,\tau,z)\vert \tau =\mp \tau_1, z = 0 \}.
\end{align*}
The surface integrals in Eq. \eqref{eqn:perp-corrDrag} are calculated using the Lorentz reciprocal theorem, similar to the single particle problem (see details in Appendix \ref{app:lorentz-perp}).
The line integrals in Eq. \eqref{eqn:perp-corrDrag}  represent the force contribution from the interfacial deformations at TCLs (see \cite{zhou2022}).

\begin{figure}
     \centering
     \includegraphics[scale=0.65]{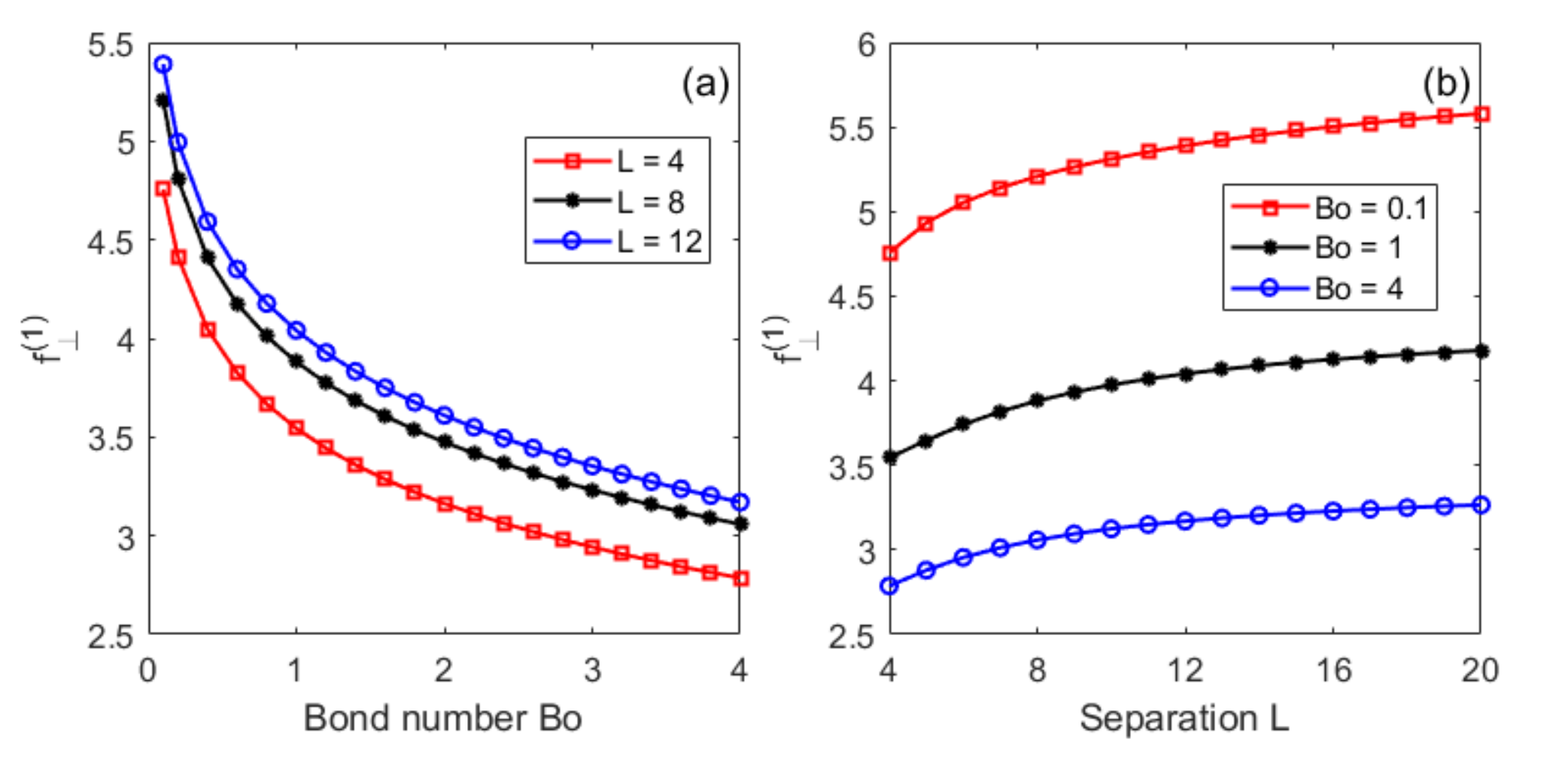}
     \caption{The drag coefficient plotted as functions of (a) Bond number Bo and (b) separation $L$. }
      \label{fig:perp-dragBoL}
     \end{figure}

Evaluating $F_\perp^{(1,0)}$ and $F_\perp^{(0,1)}$ using the trapezoidal rule, we obtain the truncated asymptotic expansion for the drag force 
\begin{align}
    F_\perp = 6\pi (\lambda+1) f_\perp^{(0)}(L) + \delta \tilde{\theta}_s (\lambda-1) f_\perp^{(1)}(\mbox{Bo},L), \label{eqn:perp-dragFormula3}
\end{align}
     where $f_\perp^{(0)}$ is the leading order drag coefficient obtained by Goldman et al. \cite{Goldman1966}, and $f_\perp^{(1)}$ is the correction drag coefficient in terms of Bo and $L$, which is shown in Fig. \ref{fig:perp-dragBoL}. Note that the flow-induced drag  $F_\perp^{(1,0)}$ integrates to zero due to anti-symmetry.  Fig. \ref{fig:perp-dragBoL}(a) shows that an increase in Bo reduces the drag coefficient $f_\perp^{(1)}$. In Fig. \ref{fig:perp-dragBoL}(b), we see that $f^{(1)}_\perp$ decreases as the separation distance $L$ decreases. This is because as the particles become closer to each other, the total amount of interfacial deformation around them decreases, and thus the correction drag caused by the deformation decreases. In the limit of large separation, the flow field and the interfacial deformation near each particle converge toward the single-particle solutions, and the correction drag coefficient $f_\perp^{(1)}/2$ converges to the value of $f^{(1)}$ in Eq.  \eqref{eqn:single-truncatedDrag}.

\subsection{Flow parallel to the particles' line-of-centers }
\label{subsect:parallel-flow}
\begin{figure}
    \centering
    \includegraphics[width=11cm]{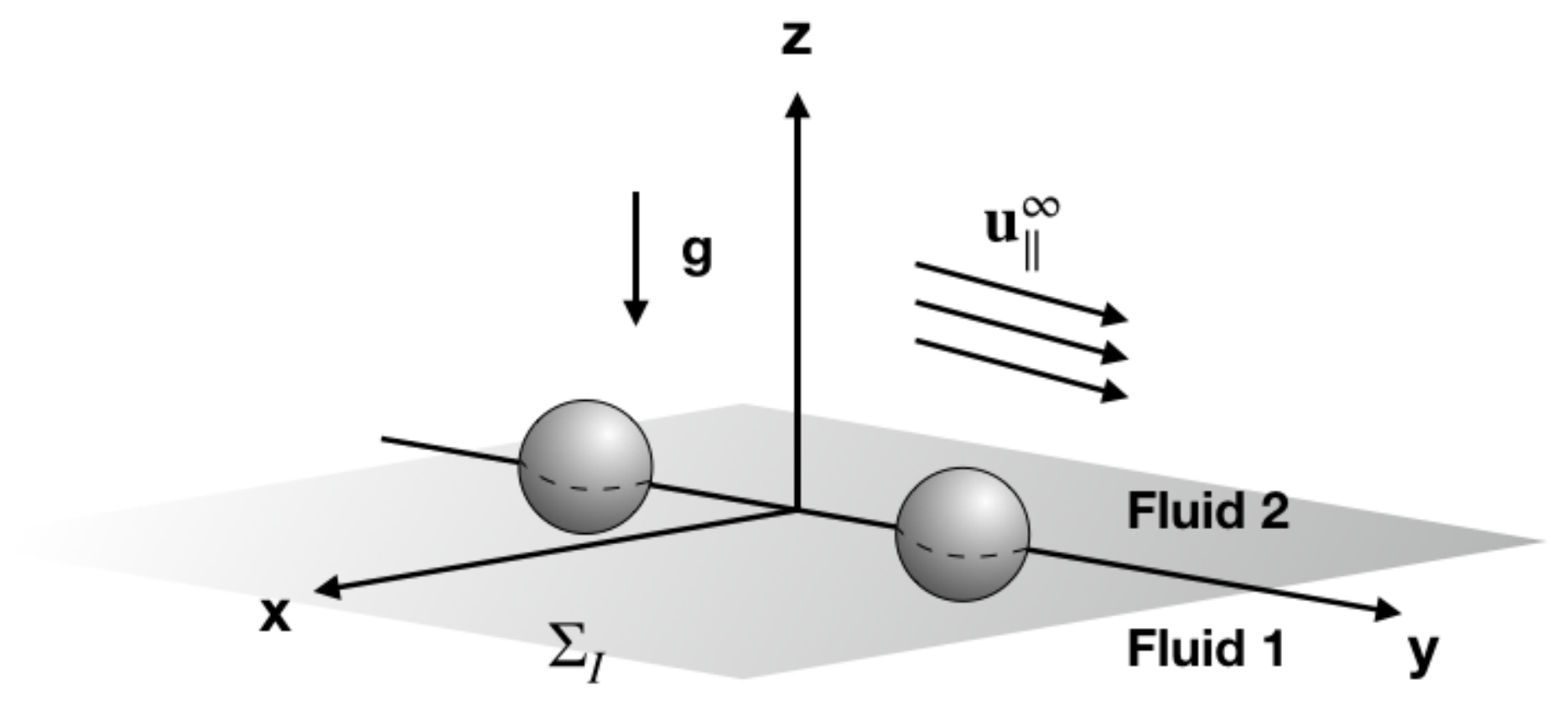}
    \caption{Illustration of two spherical particles straddling a fluid interface between two immersible viscous fluids, where the background uniform flow is parallel to the line-of-centers of the two particles.}
    \label{fig:para-setup1}
\end{figure}

Next, we consider the problem of two spherical particles at a fluid interface undergoing uniform flow parallel to their line-of-centers (see Fig. \ref{fig:para-setup1}). The centers of masses of the particles are located at $-L/2\hat{\mathbf{e}}_y$ and  $L/2\hat{\mathbf{e}}_y$, respectively, and the nondimensionalized background flow is denoted by $\mathbf{u}_\parallel ^\infty= \hat{\mathbf{e}}_y.$ We use the same asymptotic approach as we did for the previous problems and adopt similar notations.

Stimson and Jeffery \cite{Stimson1926} solved the problem of two spheres translating parallel to their line-of-centers at low Reynolds number using the stream function method, which gives us the analytical solution for the leading order velocity field $\mathbf{u}_\parallel^{(0,0)}$ (see Appendix \ref{app:para-leading-sol}). 

A similar bicylindrical coordinates system $(\sigma,\tau,z)$ is introduced: 
\begin{align}
     x = \frac{c \sin\sigma}{\cosh\tau - \cos\sigma},\quad y = \frac{c \sinh\tau}{ \cosh\tau -\cos\sigma}, \quad  -\pi < \sigma < \pi, \quad -\tau_1 < \tau < \tau_1, 
\end{align}
where $ c = \sqrt{(L/2)^2 -1}$, and $\tau = \pm \tau_1 = \pm \mbox{arccosh} (L/2)$ describe the equilibrium TCLs at the  particle surfaces.

\subsubsection{Leading order pressure recovery}
Unlike the previous cases, the jump in the pressure across the interface needed for the drag calculation is not given explicitly in \cite{Stimson1926}. But this can be found numerically by solving the differential equation for $(\lambda-1)\tilde{p}_\parallel =  p_\parallel^{2(0,0)} - p_\parallel^{1(0,0)}$, the leading order pressure difference across the flat fluid interface $\Sigma_{I}^{(0)}$. The equation that $\tilde{p}_\parallel$ satisfies is 
\begin{align}
    & \frac{\partial^2 \tilde{p}_\parallel}{\partial x^2} + \frac{\partial^2 \tilde{p}_\parallel}{\partial y^2} = -  \frac{1}{x}   \left(  \frac{\partial^2 u_{\parallel x}^{(0,0)} }{\partial x^2 } + \frac{1}{x}\frac{\partial u_x^{(0,0)} }{\partial x } + \frac{\partial^2 u_{\parallel x}^{(0,0)}}{\partial y^2} - \frac{u_{\parallel x}^{(0,0)}}{x^2}  \right),\label{eqn:para-pressureEqn4a}\\
    & \frac{\partial \tilde{p}_\parallel}{\partial n} =  \nabla^2  \mathbf{u}_\parallel^{(0,0)}\cdot \tilde{\mathbf{n}}^{(0,0)}\quad \mbox{ at } \Sigma^{(0)}_{\text{\scriptsize TLC}^{\text{\tiny I,II}}}, \label{eqn:para-pressureEqn4}\\
    & \tilde{p}_\parallel \rightarrow 0 \quad \mbox{as } \vert \mathbf{x} \vert \rightarrow \infty,
\end{align}
where $\partial/\partial n$ denotes the normal derivative at the base TCLs, and the boundary condition \eqref{eqn:para-pressureEqn4} is obtained by taking the normal component of the momentum equation (see Appendix \ref{app:pressure-recovery} for the detailed derivation of Eq. \eqref{eqn:para-pressureEqn4a}). In bicylindrical coordinates, the problem for $\tilde{p}_\parallel$ is given by
\begin{align}
 &\frac{(\cosh\tau-\cos\sigma)^2}{c^2} \left( \frac{\partial^2 \tilde{p}_\parallel}{\partial \sigma^2 } + \frac{\partial^2 \tilde{p}}{\partial \tau^2} \right) =   F(\sigma,\tau), \\
 &\left.\frac{\partial \tilde{p}_\parallel}{\partial n } \right\vert_{\tau = \pm\tau_1}= \tilde{f}(\sigma,\pm\tau_1),\\
 & \tilde{p}_\parallel(-\pi,\tau) = \tilde{p}_\parallel ( \pi,\tau)  ,\quad \frac{\partial \tilde{p}_\parallel}{\partial \sigma}  (-\pi,\tau) = \frac{\partial \tilde{p}_\parallel}{\partial \sigma} (-\pi,\tau),\\
&  \tilde{p}_\parallel(0,0) = 0 \quad ( |\mathbf{x}|\rightarrow \infty \implies (\sigma,\tau) \rightarrow (0,0)),
 \end{align}
 where 
 \begin{align}
    &  \tilde{F}(\sigma,\tau) = -  \frac{1}{x}   \left(  \frac{\partial^2 u_{\parallel x}^{(0,0)} }{\partial x^2 } + \frac{1}{x}\frac{\partial u_{\parallel x}^{(0,0)} }{\partial x } + \frac{\partial^2 u_{\parallel x}^{(0,0)}}{\partial y^2} - \frac{u_{\parallel x}^{(0,0)}}{x^2}  \right),\\
     & \tilde{f}(\sigma,\pm\tau_1) =  \nabla^2 \mathbf{u}_\parallel^{(0,0)}  \cdot \tilde{\mathbf{n}}(\sigma,\pm\tau_1) .
 \end{align}
The unit normal vector to the base TCLs, $\tilde{\mathbf{n}}^{(0,0)}$, is
 \begin{align}
      & \tilde{\mathbf{n}}^{(0,0)} \big\vert_{\tau=\pm \tau_1} = \mp \hat{\mathbf{e}}_\tau =   \mp \left(  -\frac{\sin\sigma \sinh(\pm \tau_1)}{\cosh(\pm\tau_1) - \cos\sigma} \hat{\mathbf{e}}_x + \frac{1-\cosh\sigma \cosh(\pm\tau_1)}{\cosh(\pm\tau_1) - \cos\sigma} \hat{\mathbf{e}}_y \right).
  \end{align}
This partial differential equation for $\tilde{p}_\parallel$ is solved numerically using the finite difference method and MATLAB's backslash operator to invert the discretized linear system of difference equations, and the numerical solutions show quadratic convergence \cite{zhou2022}.  

\subsection{Interfacial deformations and drag force}
The static deformation $\delta h_\parallel^{(0,1)}$, induced by the contact angle, describes the equilibrium interface shape in the absence of flow. Thus, the static deformation is unaffected by the flow orientation and $h_\perp^{(0,1)} \equiv  h_\parallel^{(0,1)}$ with the proper axis rotation. The $\mathcal{O}(\mbox{Ca})$ interfacial deformation $h^{(1,0)}_\parallel$, induced by the background flow, satisfies the stress balance equation 
\begin{align}
    \nabla h_\parallel^{(1,0)} -\mbox{Bo}h_\parallel^{(1,0)} = - \hat{\mathbf{e}}_z \cdot [\bm{\sigma}^{(0,0)}_\parallel]\cdot \hat{\mathbf{e}}_z, \label{eqn:para-deformationEqn1}
\end{align}
where the stress difference is given by 
\begin{align}
    [\bm{\sigma}^{(0,0)}_\parallel] = (\lambda-1) ( -\tilde{p}_\parallel +  \nabla[\mathbf{u}^{(0,0)}_\parallel  ] + (\nabla[\mathbf{u}^{(0,0)}_\parallel ] )^T ).
\end{align}
The stress balance equation \eqref{eqn:para-deformationEqn1} is solved as before using a second order centered finite difference method (see \cite{zhou2022} for details). 

\begin{figure}
     \centering
     \includegraphics[scale=0.65]{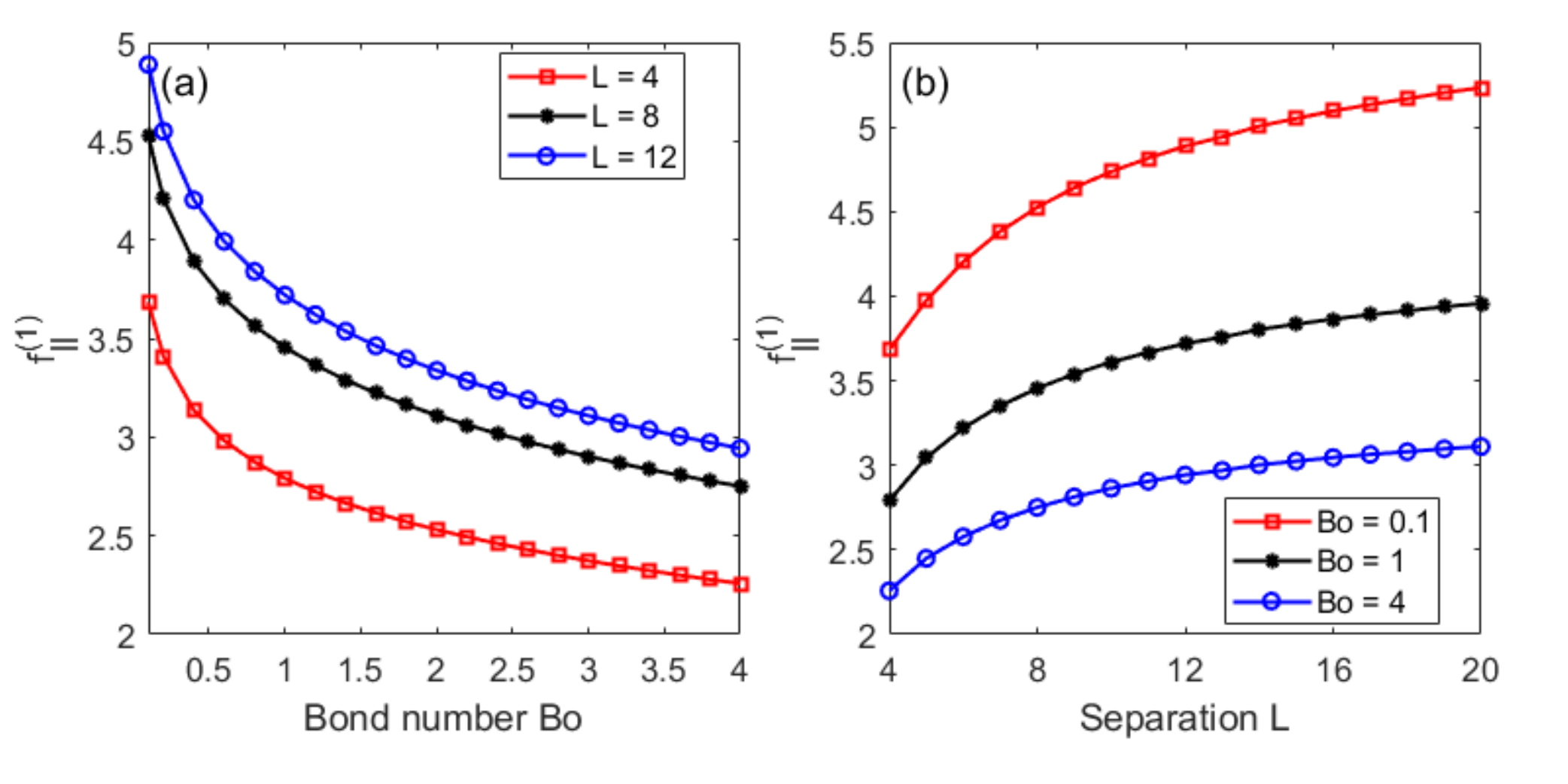}
     \caption{The drag coefficient plotted as functions of (a) Bond number Bo and (b) separation $L$. }
      \label{fig:para-dragBoL}
     \end{figure}

We use the same approach as in the previous cases to obtain the drag force exerted on the two particles, which is given in the form of a truncated asymptotic expansion: 
\begin{align}
    F_\parallel = 6\pi (\lambda+1)f_\parallel^{(0)}(L) + \delta \tilde{\theta}_s (\lambda-1)f_\parallel^{(1)} (\mbox{Bo},L), \label{eqn:para-dragFormula}
\end{align}
where $f_\parallel^{(0)}$ is the leading order drag coefficient obtained by Stimson and Jeffery \cite{Stimson1926}, and $f_\parallel^{(1)}$ is the correction drag coefficient for the contact angle induced deformation $\delta h^{(0,1)}$. The drag contribution from the flow-induced deformation $\mbox{Ca}h^{(1,0)}$ integrates to zero due to anti-symmetry. Fig. \ref{fig:para-dragBoL} shows the drag coefficient $f_\parallel^{(1)}$ as a function of Bo and $L$. The dependence of $f_\parallel^{(1)}$ on Bo and $L$ is similar to that found for the perpendicular flow past two spheres.

As the separation distance $L$ increases, the value of $f_\parallel^{(1)}/2$ converges to the single-particle drag coefficient $f^{(1)}$ in Eq. \eqref{eqn:single-truncatedDrag}. However, a slower convergence is observed compared to the case of two particles in a perpendicular flow. This can be explained by the difference in the convergence rates of the leading order solutions, i.e., $f_\parallel^{(0)} \sim 1 - 3/2L$ and $ f_\perp^{(0)} \sim 1 - 3/4L$ for $L \gg 1$ \cite{Stimson1926,Goldman1966}.

\subsection{Arbitrarily oriented flow}
\begin{figure}
    \centering
    \includegraphics[width=10cm]{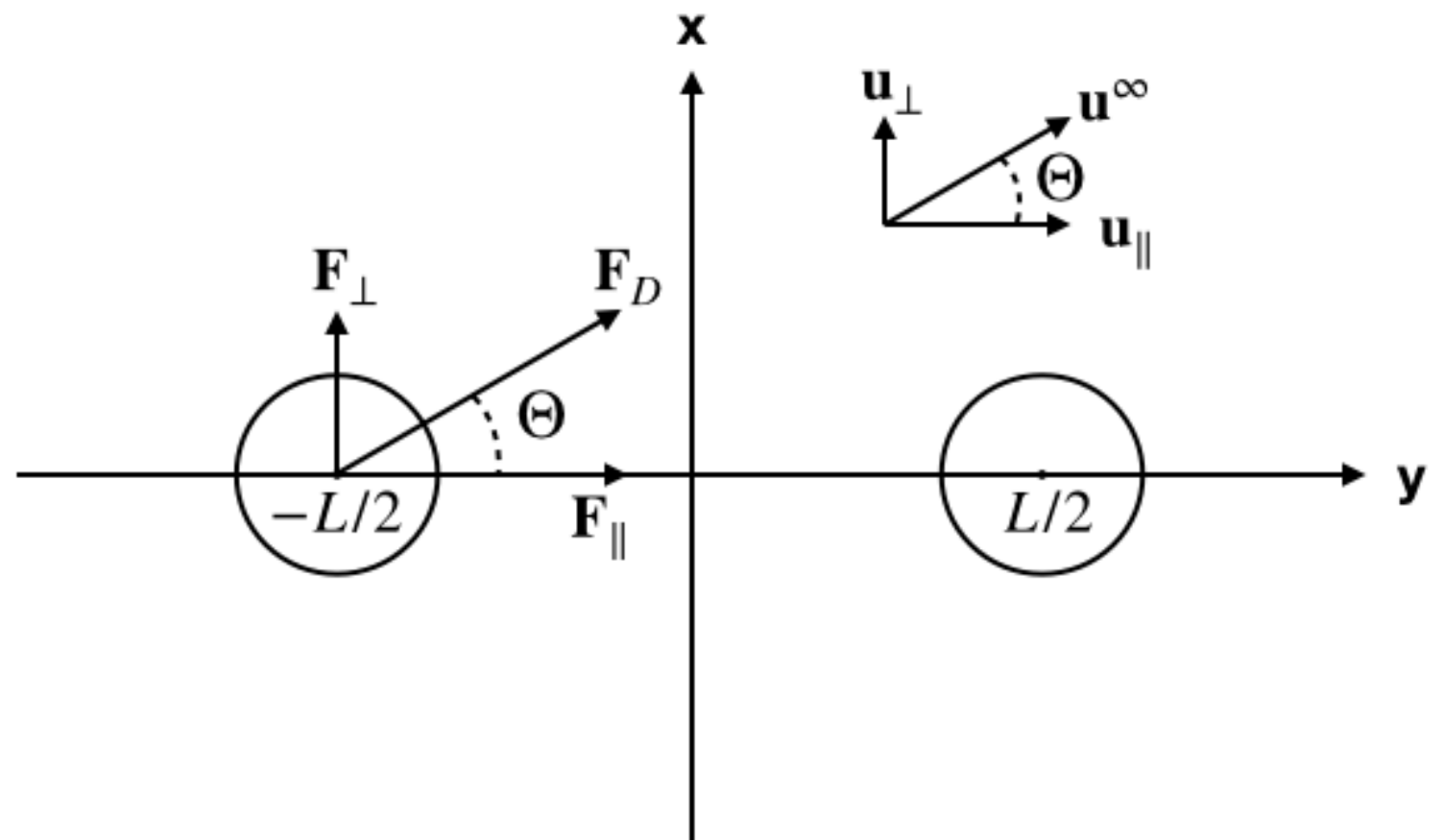}
    \caption{Sketch of the $x$-$y$ cross-section view of two particles at a fluid interface undergoing arbitrarily oriented uniform flow in the $x$-$y$ plane.}
    \label{fig:arb-setup1}
\end{figure}

\begin{figure}
    \centering
    \includegraphics[scale=0.65]{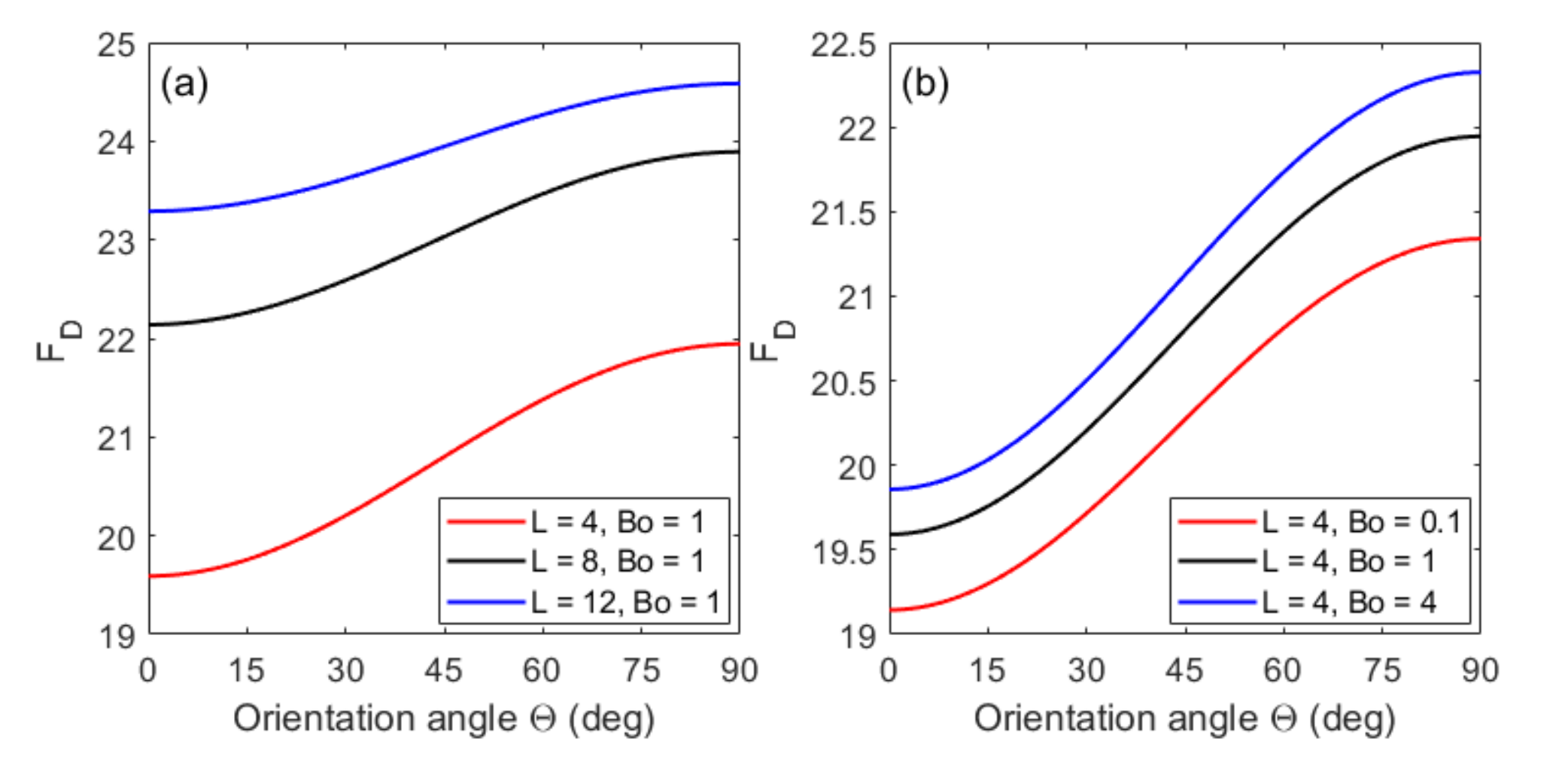}
    \caption{The magnitude of the drag force plotted as a function of the orientation angle $\Theta$ with $\delta = \theta_s = 1,  \lambda = 1/2$, and varying values of separation $L$ (a) and Bond number Bo (b). }
    \label{fig:arb-dragAngle}
\end{figure}

The analyses of two spherical particles undergoing flows perpendicular and parallel to their line-of-centers allow the calculation of arbitrarily oriented flow past two spheres at an interface. As illustrated in Fig. \ref{fig:arb-setup1}, the uniform background flow $\mathbf{u}^\infty$ is oriented at an arbitrary angle relative to the spheres' line-of-centers. The flow $\mathbf{u}^\infty$ can be decomposed into a perpendicular component to the line-of-centers and a parallel component, i.e.,
\begin{align}
    \mathbf{u}^\infty = \mathbf{u}_\perp^\infty + \mathbf{u}_\parallel^\infty \quad (|| \mathbf{u}^\infty|| = 1),
\end{align}
with 
\begin{align}
    \mathbf{u}_\perp^\infty = \sin\Theta \hat{\mathbf{e}}_x, \quad \mathbf{u}_\parallel^\infty = \cos\Theta \hat{\mathbf{e}}_y,
\end{align}
where $\Theta$ denotes the angle between the flow direction and the line-of-centers of the two spheres (see Fig. \ref{fig:arb-setup1}). 
The linearity of the Stokes equation and the boundary conditions of the $\mathcal{O}(\mbox{Ca})$ and $\mathcal{O}(\delta)$ problems allows us to calculate the drag force acting on the particles, $\mathbf{F}_D$, by vectorially combining the forces exerted by the perpendicular flow $ \mathbf{u}_\perp^\infty$ and the parallel flow $ \mathbf{u}_\parallel^\infty$, i.e.,
\begin{align}
    \mathbf{F}_D = \mathbf{F}_\perp + \mathbf{F}_\parallel
\end{align}
with \begin{align}
    \mathbf{F}_\perp = F_\perp \sin\Theta \hat{\mathbf{e}}_x,\quad \mathbf{F}_\parallel = F_\parallel \cos\Theta \hat{\mathbf{e}}_y,
\end{align}
where $F_\perp$ and $F_\parallel$ are given in Eqs. \eqref{eqn:perp-dragFormula3} and \eqref{eqn:para-dragFormula}, respectively. The magnitude of the drag force is given by 
\begin{align}
   F_D =  \vert \vert \mathbf{F}_D\vert \vert = \sqrt{ (F_\perp \sin\Theta )^2 + (F_\parallel \cos\Theta)^2}.
\end{align}
In Fig. \ref{fig:arb-dragAngle}, $F_D$ is plotted as a function of the orientation angle $\Theta.$ Given the same set of parameters, the drag force has a larger magnitude when the background flow is perpendicular to the line-of-centers than when parallel. As $\Theta$ increases from $0^\circ$ to $90^\circ$, the perpendicular component in the background flow becomes more dominant and $F_D$ increases.

\subsection{Capillary attraction force}

The dimensionless capillary force scaled by $\gamma a$ exerted on the spherical particle due to the interfacial deformation can be computed by integrating the capillary stress along the TCL, i.e., 
\begin{align}
   \mathbf{F}_C = \int_{\Sigma_\text{\scriptsize TLC}}   \tilde{\mathbf{n}}_C \mbox{ d}s,
\end{align}
where $\tilde{\mathbf{n}}_C = \tilde{\mathbf{t}}_C \times \hat{\mathbf{n}}$ is the capillarity unit vector that is normal to the TCL and lies in the interface, and $\tilde{\mathbf{t}}_C$ is the unit tangent vector to the TCL.  Let $\mathbf{r}_{C}=\hat{\mathbf{e}}_r + \hat{\mathbf{e}}_z h(r,\phi) \hat{\mathbf{e}}_z$ denote the position vector describing a point on the TCL at the particle surface.

Substituting all expansions into this vector formula,  we find that the capillarity unit vector $\tilde{\mathbf{n}}_C$ is
\begin{align}
   \tilde{\mathbf{n}}_C =&   \tilde{\mathbf{t}}_{C}\times \hat{\mathbf{n}} = \hat{\mathbf{e}}_r + \left( \mbox{Ca}\frac{\partial h^{(1,0)}}{\partial r} +  \delta \frac{\partial h^{(0,1)}}{\partial r}\right)\hat{\mathbf{e}}_z,
\end{align}
and to order Ca and $\delta$, the capillary force is
\begin{align}
    \mathbf{F}_C = F_C \hat{\mathbf{e}}_z =  & \int_0^{2\pi}   \left( \mbox{Ca}\frac{\partial h^{(1,0)}}{\partial r}(1,\phi) +  \delta \frac{\partial h^{(0,1)}}{\partial r}(1,\phi)\right)\hat{\mathbf{e}}_z  \mbox{ d}\phi \\
    = &  -2 \pi  \delta  (-\tilde{\theta}_s + \tilde{b}) \sqrt{\mbox{Bo}} C_0 K_1(\sqrt{\mbox{Bo}})\hat{\mathbf{e}}_z, \label{eqn:capillary-singleParticle}
\end{align}
where $C_0$ is defined in Eq. \eqref{eqn:single-staticDeformSol}. We observe that the flow-induced deformation does not contribute to the capillary force at leading order due to the anti-symmetric pattern of the interfaical height at the TCL. The interfacial height depends on the azimuthal angle $\phi$ in the form of $\sin\phi$ and the capillary stress along the TCL integrates to zero. The leading order contribution to $\mathbf{F}_C$ comes from the static deformation, which yields an $\mathcal{O}(\delta)$ capillary force  in the vertical ($\hat{\mathbf{e}}_z$) component. Note that the lateral capillary force is zero at leading order because the corrections to the shape of the TCL is at orders $\delta$ and Ca in the vertical component and at higher orders in the horizontal components.

\begin{figure}
    \centering
    \includegraphics[scale=0.65]{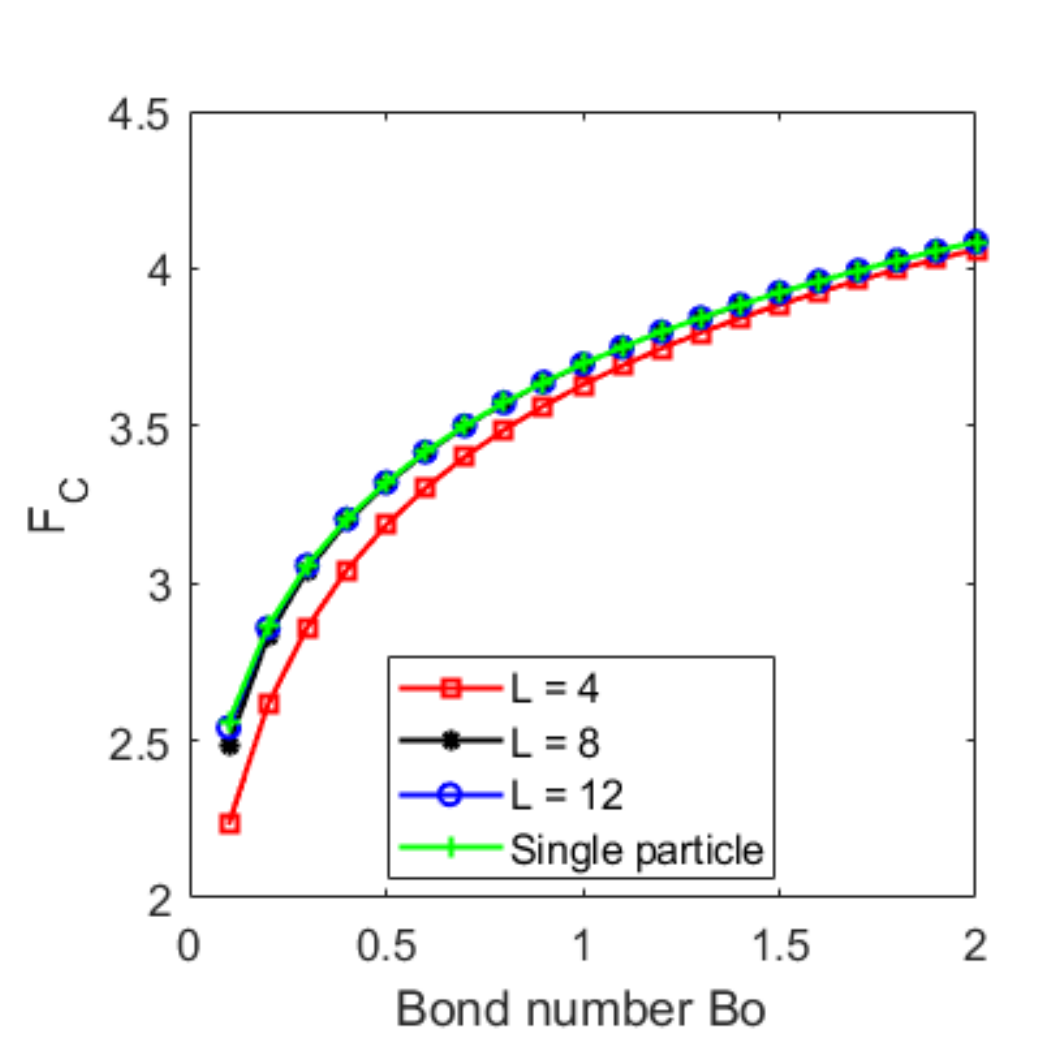}
    \caption{The dimensionless capillary force plotted as a function of the Bond number Bo. The green curve shows the single-particle capillary force; the red, black and blue curves are the capillary forces exerted on one of the two particles due to the two-particle deformations in perpendicular background flows, with separation distance 
    $L = 4, 8$ and 12, respectively. Parameter values: $\mbox{Ca}=1,\delta=\tilde\theta_s =1, \tilde{b}=0$.  }
    \label{fig:capillary-perp}
\end{figure}

Paralleling the single-particle capillary force calculation, we are able to obtain capillary forces exerted by the deformed interface near two spherical particles. We first consider the case where the two particles at the interface undergo uniform flows perpendicular to their line-of-centers (see Appendix \ref{app:capillaryForce} for detailed calculations). 
The leading order capillary force $F_C$ is given by Eq. \eqref{eqn:app-F_C}. Similar to the single-particle problem, the flow-induced deformation does not contribute to the capillary force at leading order due to the anti-symmetry argument. Also, due to the symmetric static interface shape in the direction of the line-of-centers of the particles, the vertical capillary forces exerted on the two particles are identical.  In Fig. \ref{fig:capillary-perp}, the capillary forces due to the single-particle deformation and the two-particle deformations with perpendicular background flows are plotted as a function of the Bond number. As the separation distance increases, the static deformation near each particle converges to the single-particle deformation, and we see that the capillary force exerted on each particle converges to the single-particle capillary force.   

\begin{figure}
    \centering
    \includegraphics[scale=0.65]{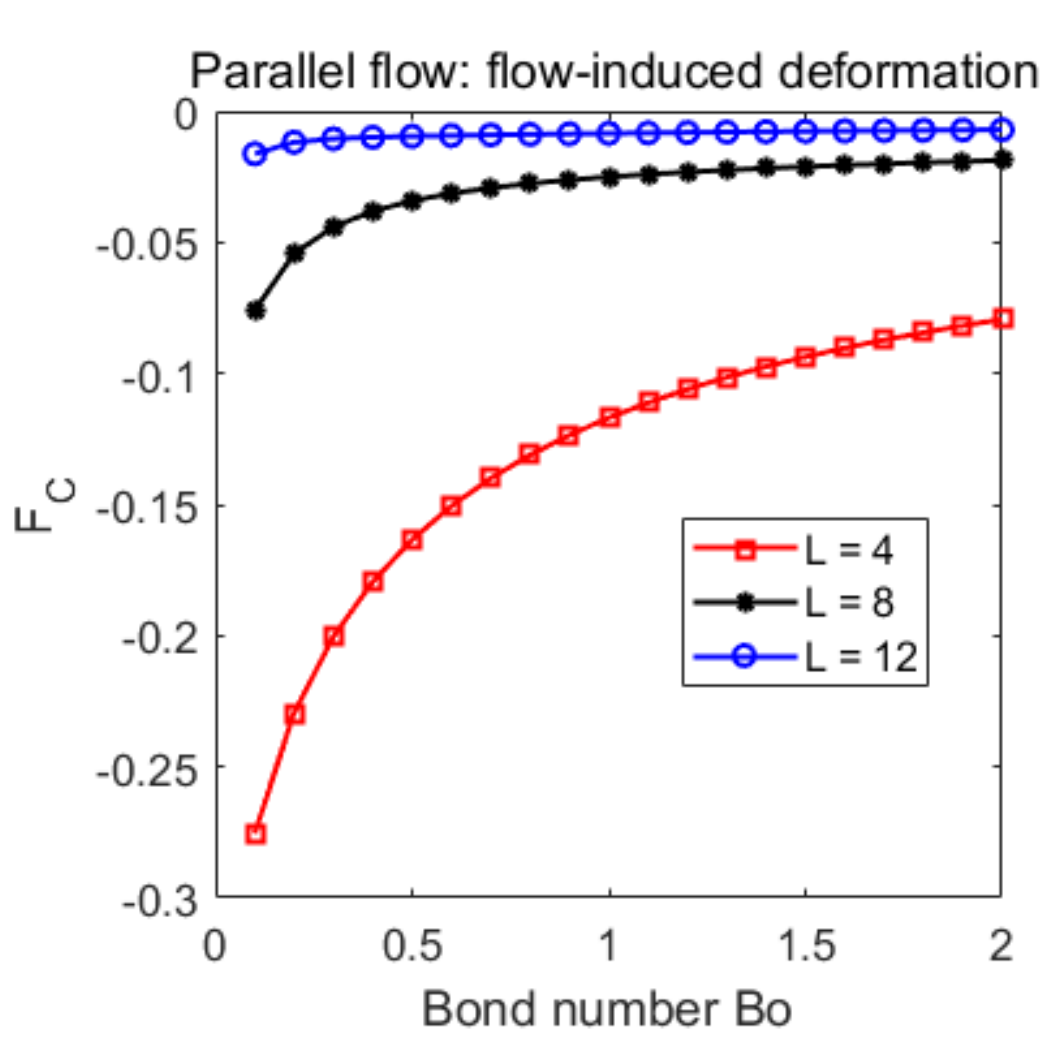}
    \caption{The dimensionless capillary force exerted on particle I ($\Sigma_{P}^\text{\scriptsize I }$) due to the parallel flow-induced deformations plotted as a function of the Bond number Bo, with separation  $L = 4, 8,$ and 12 (Ca $=1,\lambda=0$).}
    \label{fig:capillary-para-flowInducedDeformation}
\end{figure}

\begin{figure}
    \centering
    \includegraphics[scale=0.65]{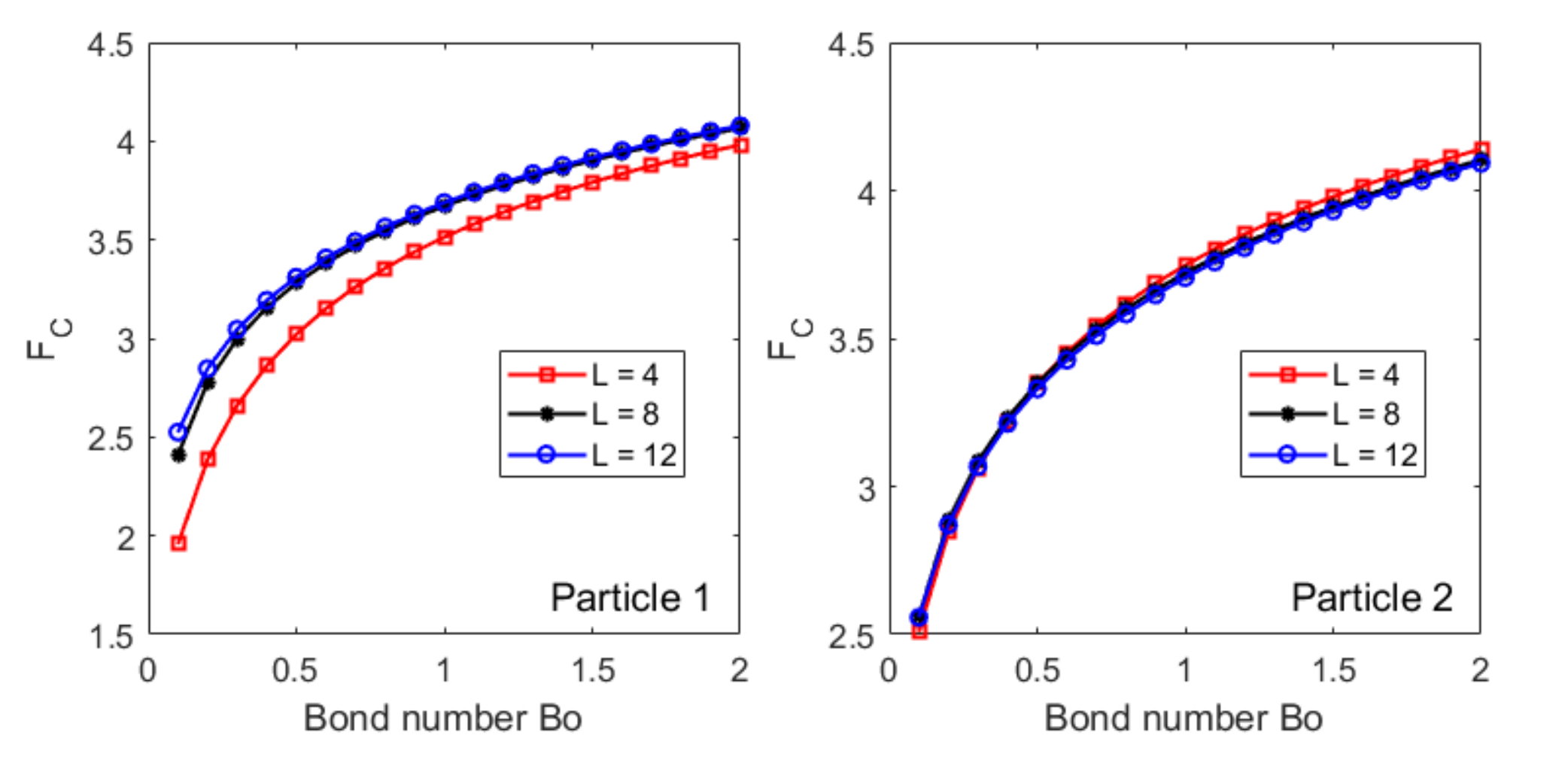}
    \caption{The dimensionless capillary force due to the total deformations near two particles undergoing parallel background flows. Parameter values: $\mbox{Ca}=1,\delta=\tilde\theta_s =1,\lambda=0. $ }
    \label{fig:capillary-para-totalDeformations}
\end{figure}

In the case of two particles at an interface in a parallel background flow, the flow-induced deformation's contribution to the leading order capillary force does not vanish because the two particles' interaction with the flow breaks the anti-symmetry at the TLCs. As a result, the flow-induced deformations at the two particles' TCLs have the same magnitude but opposite signs, which implies the $\mathcal{O}(\mbox{Ca})$ capillary forces acting on the two particles also have the same magnitude and different signs. Fig. \ref{fig:capillary-para-flowInducedDeformation} shows the capillary force exerted on the particle centered at $-L/2\hat{\mathbf{e}}_y$ ($\Sigma_P^\text{\scriptsize I}$) as a function of the Bond number Bo, with different values of separation $L.$ 
Since the orientation of the background flow does not affect the static deformation, the $\mathcal{O}(\delta)$ capillary force is identical to the one in the perpendicular flow problem. In Fig. \ref{fig:capillary-para-totalDeformations}, we show the leading order capillary force acting on each particle due to the total deformations. 

The lateral capillary force at higher orders was calculated by Vella and Mahadevan \cite{Vella2005}, where they derived the formula of the (dimensionless) capillary force in the absence of flow from the Nicolson approximation \cite{Chan1981, Kralchevsky1994, Nicolson1949}:
\begin{align}
    \tilde{F}_C = 2\pi \mbox{Bo}^{5/2} \Sigma^2 K_1(\sqrt{\mbox{Bo}}L),
\end{align}
where $  \Sigma = \frac{2(\rho_P-\Delta \rho) - 1}{3} -\frac{1}{2}\cos\theta_s + \frac{1}{6}\cos^2\theta_s$ and $\rho_P$ is the particle density. The assumptions that the contact angle is close to $90^\circ$ and the immersion depth is small ($\theta_s = \pi/2+\delta\tilde\theta_s , b = \delta\tilde{b}$) imply $\Sigma \sim \delta $ and $ \tilde{F}_C \sim \delta^2$, which verifies our discovery that the static deformation does not contribute to the lateral capillary force at order $\delta.$ D\"orr and Hardt \cite{dorr2015} studied the pair interaction of particles by constructing the interfacial deformation around two particles via linear superposition of the single-particle deformation. Under the assumptions of rotated and pinned TCLs and a large particle separation, the interfacial deformation is induced solely by the uniform background flow. Similar to our analysis,  D\"orr and Hardt's calculation shows that the leading order capillary force is in the vertical direction and the lateral capillary force comes  at higher orders and that the vertical capillary force vanishes when the uniform background flow is perpendicular to the particles' line-of-centers.

\section{Conclusions}
In this work, we have studied the problems of fluid motion past one and two spherical particles attached to a deformable fluid interface undergoing uniform Stokes flow. Using the two-parameter asymptotic expansions for small Capillary number and correction contact angle, we have obtained the analytical expressions for the flow-induced deformation and the static deformation (induced by the contact angle) around a single particle. In the two-particle problems, where the background flow is perpendicular or parallel to the particles' line-of-centers, similar deformation solutions were calculated numerically using finite difference methods.   
To study the effects of interfacial deformations on the drag force exerted on the particles, we used the Lorentz reciprocal theorem to derive analytical expressions for the correction drag forces in terms of the zeroth-order approximations and the deformation solutions. For the single-particle problem, the drag force is given in the form in Eq. \eqref{eqn:single-truncatedDrag}, where the drag caused by the flow-induced deformation integrates to zero due to its anti-symmetric configuration in the flow direction, and the correction drag caused by the static deformation is shown to linearly depend on the correction contact angle and the viscosity difference and have a nonlinear dependence on the Bond number. The Bond number characterizes the density mismatch between the two fluid phases, and an increase in the density mismatch flattens the interface shape near the particle, which reduces the effect of the interfacial deformation on the drag force. The normalized drag $F_D^*$ (see Eq. \eqref{eqn:single-normalizedDrag}) is shown to be consistent with the 2D numerical results by Loudet et al. \cite{Loudet2020}.

For the two-particle problems, we derived the first-order approximations from the solutions of two spheres translating perpendicular and parallel to their line-of-centers in a viscous fluid
\cite{Goldman1966,Stimson1926}. Similar to the single-particle problem, the flow-induced interfacial deformations do not affect the drag force acting on the particles at leading order. The corrected drag forces for the static deformations are given by Eqs. \eqref{eqn:perp-dragFormula3} and \eqref{eqn:para-dragFormula}. A more general solution where the uniform background flow is arbitrarily oriented relative the particles' line-of-centers has been obtained by vectorially combining the drag forces exerted by the perpendicular and parallel flows.  Our predictions for the drag also compares well with the experimental results of Petkov et al. \cite{Petkov1995} (see Appendix \ref{app:petkov}). This is also true of D\"orr et al. \cite{dorr2016} 's model which has different assumptions.   Additional work is needed to clarify these predictions.

In addition, we were able to calculate the capillary force exerted on the particles due to the interfacial deformation. It is shown that the static deformation contributes to the capillary force at order $\delta$ in the vertical ($\hat{\mathbf{e}}_z$) component, and the flow-induced deformation doesn't contribute at order Ca in the single-particle case and the two-particle case when the background flow is perpendicular to the particles' line-of-centers. In the case of two particles at an interface in a parallel flow, the flow-induced deformation is shown to have a nonzero contribution to the $\mathcal{O}(\mbox{Ca})$ vertical capillary force.

\begin{acknowledgments}
This work was partially supported by NSF grants DMS 1718114 and DMS 2108502. 
\end{acknowledgments}

\appendix
\section{Flow motion past a single particle}

\subsection{Leading order and correction problems}
\label{app:single-leading-corr-problems}
The $\mathcal{O}(1)$ pressure and velocity field satisfy the Stokes equations
\begin{align}
    -\nabla p^{(0,0)} + \lambda_i \nabla^2 \mathbf{u}^{(0,0)} = &  \mathbf{0} \label{eqn:app-leading-sol1}\\ 
     \nabla\cdot \mathbf{u}^{(0,0)} = &  0, 
\end{align}
 with boundary conditions 
\begin{align}
  &  \mathbf{u}^{(0,0)}(x,y,z) \rightarrow \mathbf{u}^\infty =\hat{\mathbf{e}}_y \quad \mbox{ as }\vert \mathbf{x}\vert \rightarrow \infty\\
    & \mathbf{u}^{(0,0)}(x,y,z) = \mathbf{0} \quad \mbox{ for } \mathbf{x} \in \Sigma_P^{(0)} = \{ (x,y,z) \vert x^2 + y^2 + z^2 =1 \}
\end{align}
and interface conditions 
\begin{align}
    & \mathbf{u}^{(0,0)} \cdot \hat{\mathbf{e}}_z = 0,\\
    &[ \mathbf{u}^{(0,0)}]\cdot \hat{\mathbf{t}}^{(0,0)} = 0, \\
      &\hat{\mathbf{t}}^{(0,0)}\cdot [\bm{\sigma}^{(0,0)}] \cdot \hat{\mathbf{e}}_z = 0, \label{eqn:app-leading-sol2}
\end{align}
for $\mathbf{x} = (x,y,0)$.

The $\mathcal{O}(\mbox{Ca})$ pressure and velocity field satisfy the Stokes equation 
\begin{align}
      &  -\nabla p^{(1,0)} + \lambda_i \nabla^2 \mathbf{u}^{(1,0)} = 0,\\
      & \nabla \cdot \mathbf{u}^{(1,0)}  = 0, 
\end{align}
with boundary conditions 
\begin{align}
      &  \mathbf{u}^{(1,0)}(x,y,z) \rightarrow \mathbf{0} \quad \mbox{ as }\vert \mathbf{x}\vert \rightarrow \infty\\
    & \mathbf{u}^{(1,0)}(x,y,z) = \mathbf{0} \quad \mbox{ for } \mathbf{x} \in \Sigma_P^{(0)} = \{ (x,y,z) \vert x^2 + y^2 + z^2 =1 \}
\end{align}
and interface conditions 
\begin{align}
      &  \mathbf{u}^{(1,0)}\cdot \hat{\mathbf{e}}_z = - \mathbf{u}^{(0,0)} \cdot \hat{\mathbf{n}}^{(1,0)}  - \frac{\partial \mathbf{u}^{(0,0)}}{\partial z } h^{(1)}\cdot\hat{\mathbf{e}}_z \\
    & [\mathbf{u}^{(1,0)}]\cdot\hat{\mathbf{t}}^{(0,0)} = - [\mathbf{u}^{(0,0)}] \cdot \hat{\mathbf{t}}^{(1,0)} -  \frac{\partial [\mathbf{u}^{(0,0)}]}{\partial z }  h^{(1,0)}\cdot\hat{\mathbf{t}}^{(0,0)} =0, 
    \end{align}
for $\mathbf{x}= (x,y,0).$
The correction deformation $h^{(1,0)}$ satisfies the normal stress balance equation 
\begin{align}
    -\nabla^2 h^{(1,0)} +\mbox{Bo} h^{(1,0)} = \hat{\mathbf{e}}_z \cdot [\bm{\sigma}^{(0,0)}] \cdot \hat{\mathbf{e}}_z, 
\end{align}
and the correction tangential stress balance equation is given by 
\begin{align}
    &  \hat{\mathbf{t}}^{(0,0)} \cdot  [\bm{\sigma}^{(1,0)}] \cdot \hat{\mathbf{e}}_z =  - \hat{\mathbf{t}}^{(0,0)} \cdot  \frac{\partial [\bm{\sigma}^{(0,0)}]}{\partial z }h^{(1,0)} \cdot \hat{\mathbf{e}}_z - \hat{\mathbf{t}}^{(0,0)} \cdot [\bm{\sigma}^{(0,0)}] \cdot \hat{\mathbf{n}}^{(1,0)} - \hat{\mathbf{t}}^{(1,0)} \cdot [\bm{\sigma}^{(0,0)}] \cdot \hat{\mathbf{e}}_z.
\end{align}

The $\mathcal{O}(\delta)$ pressure and velocity field satisfy the Stokes equation 
\begin{align}
      &  -\nabla p^{(0,1)} + \lambda_i \nabla^2 \mathbf{u}^{(0,1)} = 0,\\
      & \nabla \cdot \mathbf{u}^{(0,1)}  = 0, 
\end{align}
with boundary conditions 
\begin{align}
      &  \mathbf{u}^{(0,1)}(x,y,z) \rightarrow \mathbf{0} \quad \mbox{ as }\vert \mathbf{x}\vert \rightarrow \infty\\
    & \mathbf{u}^{(0,1)}(x,y,z) = -\Tilde{b} \frac{\partial\mathbf{u}^{(0,1)}}{\partial z} \quad \mbox{ for } \mathbf{x} \in \Sigma_P^{(0)} = \{ (x,y,z) \vert x^2 + y^2 + z^2 =1 \}
\end{align}
and interface conditions 
\begin{align}
      &  \mathbf{u}^{(0,1)}\cdot \hat{\mathbf{e}}_z = - \mathbf{u}^{(0,0)} \cdot \hat{\mathbf{n}}^{(0,1)}  - \frac{\partial \mathbf{u}^{i(0,0)}}{\partial z } h^{(0,1)}\cdot\hat{\mathbf{e}}_z \\
    & [\mathbf{u}^{(0,1)}]\cdot\hat{\mathbf{t}}^{(0,0)} = - [\mathbf{u}^{(0,0)}] \cdot \hat{\mathbf{t}}^{(0,1)} -  \frac{\partial [\mathbf{u}^{(0,0)}]}{\partial z }  h^{(0,1)}\cdot\hat{\mathbf{t}}^{(0,0)} =0, 
    \end{align}
for $\mathbf{x}= (x,y,0).$
The correction deformation $h^{(0,1)}$ satisfies the normal stress balance equation 
\begin{align}
    -\nabla^2 h^{(0,1)} +\mbox{Bo} h^{(0,1)} = 0, 
\end{align}
and the correction tangential stress balance equation is given by 
\begin{align}
    &  \hat{\mathbf{t}}^{(0,0)} \cdot  [\bm{\sigma}^{(0,1)}] \cdot \hat{\mathbf{e}}_z =  - \hat{\mathbf{t}}^{(0,0)} \cdot  \frac{\partial [\bm{\sigma}^{(0,0)}]}{\partial z }h^{(0,1)} \cdot \hat{\mathbf{e}}_z - \hat{\mathbf{t}}^{(0,0)} \cdot [\bm{\sigma}^{(0,0)}] \cdot \hat{\mathbf{n}}^{(0,1)} - \hat{\mathbf{t}}^{(0,1)} \cdot [\bm{\sigma}^{(0,0)}] \cdot \hat{\mathbf{e}}_z.
\end{align}

\subsection{The leading order velocity field }
In Cartesian coordinates, the leading-order velocity field given by Eqs. \eqref{eqn:app-leading-sol1}-\eqref{eqn:app-leading-sol2} is 
\begin{align}
    u_x^{(0,0)} = &  -\frac{3 x y (x^2 + y^2 + z^2 -1) }{4 (x^2 + y^2 + z^2 )^{5/2}},\label{eqn:app-single-leading-ur}\\
    u_y^{(0,0)} =  & \frac{1}{4}\left(4 - \frac{3 y^2 (x^2 + y^2 + z^2 -1)}{(x^2 + y^2 + z^2 )^{5/2}} - \frac{1}{(x^2 + y^2 + z^2 )^{3/2}} - \frac{3}{(x^2 + y^2 + z^2)^{1/2}}\right),\label{eqn:app-single-leading-uphi}\\
    u_z^{(0,0)} = & -\frac{3 yz (x^2 + y^2 + z^2 - 1)}{4 (x^2 + y^2 + z^2)^{5/2}},\label{eqn:app-single-leading-uz}
\end{align}
where $\mathbf{u}^{(0,0)} = u_x^{(0,0)} \hat{\mathbf{e}}_x + u_y^{(0)}\hat{\mathbf{e}}_y+ u_z^{(0,0)}\hat{\mathbf{e}}_z$. In cylindrical coordinates, the leading order velocity field is given by 
\begin{align}
u_r^{(0,0)} = & \frac{1}{4}\left( 4+\frac{-6r^4-z^2-3z^4+r^2(2-9z^2)}{(r^2+z^2)^{5/2}} \right)\sin\phi\\
u_\phi^{(0,0)} = & \frac{1}{4}\left( 4-\frac{1}{(r^2+z^2)^{3/2}} - \frac{3}{\sqrt{r^2+z^2}}\right)\cos\phi\\
u_z^{(0,0)} = &  - \frac{1}{4}\frac{3 r z(r^2+z^2-1)}{(r^2+z^2)^{5/2}} \sin\phi,
\end{align}
where $\mathbf{u}^{(0,0)} = u_r^{(0,0)} \hat{\mathbf{e}}_r + u_\phi^{(0,0)} \hat{\mathbf{e}}_\phi + u_z^{(0,0)} \hat{\mathbf{e}}_z $. 
  
\subsection{Comparison with experimental data}
\label{app:petkov}

\begin{figure}
    \centering
    \includegraphics[scale=0.65]{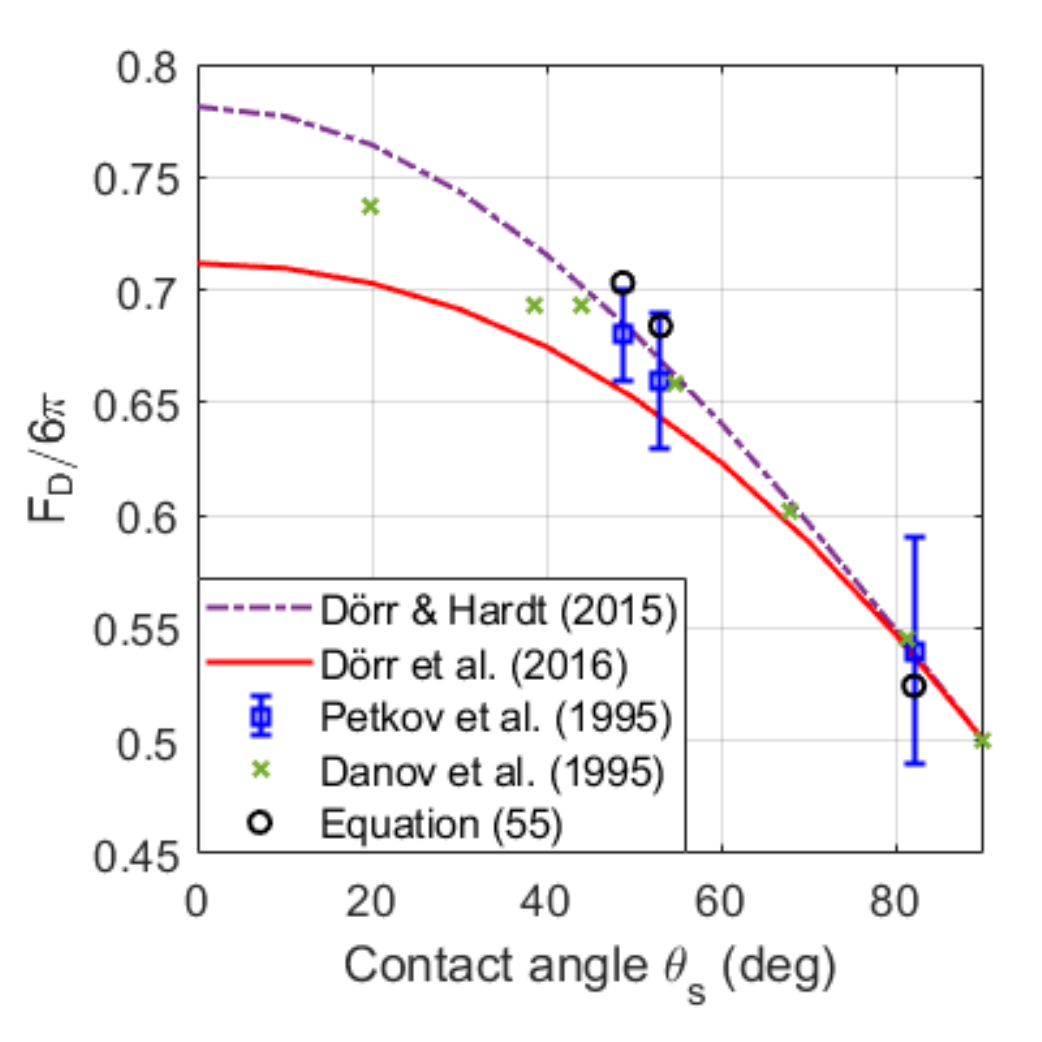}
    \caption{Comparison of the normalized drag computed from Eq. \eqref{eqn:single-truncatedDrag} with experimental data from Petkov et al. \cite{Petkov1995} and theoretical predictions from the literature \cite{Danov1995,dorr2015,dorr2016}.  }
    \label{fig:single-petkov}
\end{figure}

In the experiments by Petkov et al., the drag coefficients of spherical particles at an air-water interface with viscosity ratio $\mu_2/\mu_1\approx 0.02$ were measured at $\theta_s = 47.8^\circ, 53.0^\circ,$ and $82.0^\circ$  \cite{Petkov1995}. Petkov et al. directly measured the immersion depth and calculated the values of the contact angle via balancing vertical forces at equilibrium. Fig. \ref{fig:single-petkov} compares the result \eqref{eqn:single-truncatedDrag} with experimental and theoretical drag coefficients, where the theoretical models by \cite{Danov1995,dorr2015,dorr2016} are under the assumptions that $\mu_2/\mu_1 = 0$ and the interface remains flat. The result \eqref{eqn:single-truncatedDrag} is in good agreement with the experimental data for $\theta_s = 82.0^\circ$. Further verification is required. 

\section{Pair interaction: perpendicular flow}
\subsection{Leading order pressure and velocity field} \label{app:perp-leading-sol}
The leading order pressure $p^{(0,0)}_\perp$ and velocity field $\mathbf{u}^{(0,0)}_\perp$ obtained from \cite{Goldman1966} are given in cylindrical coordinates $(x,\rho,\varphi)$ and bicylindrical coordinates $(\eta,\xi,\phi)$ defined by 
$$
    y = \rho\cos\varphi,\quad z = \rho\sin\varphi,
$$
and 
$$
    x = \frac{c \sinh\xi}{\cosh\xi-\cos\eta}, \quad \rho = \frac{c\sin\eta}{\cosh\xi-\cos\eta},
$$
where $c = \sqrt{(L/2)^2-1}$.
The pressure and the components of the velocity field are given in terms of the auxiliary functions, i.e.,
\begin{align}
   &  p^{(0,0)}_\perp  = -\frac{\lambda_i}{c}  W \cos\phi,\\
    & u^{(0,0)}_{\perp x} =  \frac{1}{2c} ( x W +2c Z) \cos\phi  ,\\
 & u^{(0,0)}_{\perp \rho}  = -\left(  \frac{1}{2c} (\rho W + c(X+Y))  -1 \right)\cos\phi  \\
 & u^{(0,0)}_{\perp \varphi} = -\left(\frac{1}{2c} (X-Y) +1 \right)\sin\phi,
\end{align}
where $ \mathbf{u}^{(0,0)}_\perp = u^{(0,0)}_{\perp x} \hat{\mathbf{e}}_x +  u^{(0,0)}_{\perp \rho} \hat{\mathbf{e}}_\rho + u^{(0,0)}_{\perp \varphi} \hat{\mathbf{e}}_\varphi. $ 
The auxiliary functions are given in the forms:
\begin{align}
    Z = &  (\cosh\xi - \cos\eta)^{1/2} \sin\eta \sum_{n=1}^\infty A_n \sinh(n+1/2)\xi  P_n'(\cos\eta),\\
        W = &  (\cosh\xi - \cos\eta)^{1/2} \sin\eta \sum_{n=1}^\infty  B_n \cosh(n+1/2)\xi  P_n'(\cos\eta),\\
        Y = &  (\cosh\xi - \cos\eta)^{1/2}  \sum_{n=0}^\infty  D_n \cosh(n+1/2)\xi  P_n (\cos\eta),\\
                X = &  (\cosh\xi - \cos\eta)^{1/2} \sin^2\eta \sum_{n=2}^\infty  F_n \cosh(n+1/2)\xi P_n''(\cos\eta),
\end{align}
where $P_n$ denotes the Legendre polynomial of order $n.$ The formula for the coefficients $B_n, D_n,$ and $F_n $ are
\begin{align}
       B_n = &  \left[ 2 \frac{n -1}{2n-1} (\gamma_n-1)\right]A_{n-1} -2  \gamma_n A_n + \left[ 2\frac{n+2}{2n+3}(\gamma_n+1) \right]A_{n+1}  ,\\
       \begin{split}
 D_n = & 2 \sqrt{2} e^{-(n+1/2)\tau_1}\mbox{sech} (n+1/2)\tau_1 - \left[ \frac{n(n-1)}{2n-1}(\gamma_n-1) \right]A_{n-1} \\
 & +  \left[ \frac{(n+1)(n+1)}{2n+3}(\gamma_n+1)\right]A_{n+1} \end{split} \\
 F_n =  & \frac{\gamma_n-1}{2n-1} A_{n-1} - \frac{\gamma_n+1}{2n+3} A_{n+1},
\end{align}
with $\tau_1= \mbox{arccosh}(L/2)$ and  $\gamma_n = \coth \tau_1 \tanh(n+1/2)\tau_1,$ and  $A_n$ satisfies the recurrence equation 
\begin{align}
\begin{split}
  &  \left[(n-1)(\gamma_{n-1}-1) - \frac{(n-1)(2n-3)}{(2n-1)}(\gamma_n-1) \right]A_{n-1} \\
   & + \left[ (2n+1) - 5\gamma_n - \frac{n(2n-1)}{(2n+1)}(\gamma_{n-1}+1) + \frac{(n+1)(2n+3)}{(2n+1)}(\gamma_{n+1}-1) \right]A_n\\
 &   + \left[ \frac{(n+2)(2n+5)}{(2n+3)} (\gamma_n+1) -(n+2)(\gamma_{n+1}+1) \right] A_{n+1} \\
 & = \sqrt{(2)}e^{-(n+1/2)\tau_1} \left[ \frac{e^{\tau_1}}{\cosh(n-1/2)\tau_1} - \frac{2}{\cosh(n+1/2)\tau_1} +\frac{e^{-\tau_1}}{\cosh(n+3/2)\tau_1} \right]
   \end{split} \label{eqn:app-perp-recur-form-An}
\end{align}
for $n =1,2,\cdots.$ The approximate values of $A_n$, $n = 1,\cdots, N$, can be obtained by numerically solving the linear system consisting of the first $N$ equations for Eq. \eqref{eqn:app-perp-recur-form-An}.

\subsection{Contact angle conditions}
\label{app:perp-contact-angle-cond}
Inserting the asymptotic expansions \eqref{eqn:perp-asym-exp} into the LHS of Eq. \eqref{eqn:perp-contact-angle1}, we obtain 
\begin{align}
\cos(\pi/2 -\Psi_c) = &   \sin\Psi_c =  \sin( \arcsin(r_c) - \theta_s )  \label{eqn:app-perp-contact-cond-lhs}\\
     = &   \sin\left(\arcsin(\sqrt{ 1- (\mbox{Ca} h_\perp^{(1,0)}  + \delta h_\perp^{(0,1)})^2 } ) - (\pi/2 + \delta \tilde{\theta}_s)\right) \\
    =  &  - \mbox{Ca} h^{(1,0)}  + \delta \left( -h_\perp^{(0,1) } -\tilde{\theta}_s\right) + \cdots,
\end{align}
where $r_c$ denotes the radius of the TLC (see Fig. \ref{fig:perp-contact-angle}). 
Using the tools of differential geometry, the RHS of Eq. \eqref{eqn:perp-contact-angle1} can be written as (see Refs. \cite{Kralchevsky1992,McConnell1957}) 
\begin{align}
     \hat{\mathbf{e}}_\tau \cdot \hat{\mathbf{n}} \big\vert_{\tau=\pm\tau_1} = \pm \frac{1}{\sqrt{g}} \sqrt{\frac{g^*}{a^{*}_1}} \frac{\partial h_{\perp}}{\partial \tau}  \bigg\vert_{\tau = \pm\tau_1}, \label{eqn:app-perp-contact-cond-rhs}
\end{align}
where 
$
    g = c^2/(\cosh\tau - \cos\sigma)^2 
$
is the component of the metric tensor of the bipolar coordinate system given in Eq. \eqref{eqn:perp-bicylindricalCoordinates}, $g^* = g^2$ is the determinant of the metric tensor, and $a^{*}$ is the determinant of the surface metric tensors of $z = h_\perp(\sigma,\tau)$, which is defined by 
\begin{align}
     a^* = g^2 \left[ 1 + \frac{1}{g}  \left(\frac{\partial h_{\perp}}{\partial \sigma}  \right)^2 \right]  \left[   1 + \frac{1}{g}  \left(\frac{\partial h_{\perp}}{\partial \tau} \right)^2 \right].\label{eqn:app-perp-contact-cond-a}
\end{align}
Substituting the expansion of $h_\perp$ into Eq. \eqref{eqn:app-perp-contact-cond-a} yields
\begin{align}
     a^*  = g^2 + & \mbox{Ca}^2 g \left[\left(\frac{\partial h_{\perp}^{(1,0)}}{\partial \tau}  \right)^2 + \left(\frac{\partial h_{\perp}^{(1,0)}}{\partial \sigma}  \right)^2  \right]  + \delta^2 g \left[\left(\frac{\partial h_{\perp}^{(0,1)}}{\partial \tau}  \right)^2 + \left(\frac{\partial h_{\perp}^{(0,1)}}{\partial \sigma}  \right)^2    \right] \\
    &   + 2 \mbox{Ca}\delta g \left[\frac{\partial h_{\perp}^{(1,0)}}{\partial \tau}\frac{\partial h_{\perp}^{(0,1)}}{\partial \tau} + \frac{\partial h_{\perp}^{(1,0)}}{\partial \sigma}\frac{\partial h_{\perp}^{(0,1)}}{\partial \sigma} \right] + \cdots.
\end{align}
We assume the separation coefficient $c$ is of finite order, i.e., $c^2 =\mathcal{O}(1).$ Then, $g = \mathcal{O}(1)$ and  $a^* \approx g^2$. Combining Eqs. \eqref{eqn:app-perp-contact-cond-lhs} - \eqref{eqn:app-perp-contact-cond-rhs} and collecting coefficients of $\mbox{Ca}$ and $\delta$, we obtain 
\begin{align}
   &  \pm \frac{\cosh\tau-\cos\sigma}{c}\frac{\partial h_\perp^{(1,0)}}{\partial \tau} + h_\perp^{(1,0)} = 0,\\
      &  \pm \frac{\cosh\tau-\cos\sigma}{c}\frac{\partial h_\perp^{(0,1)}}{\partial \tau} + h_\perp^{(0,1)} =- \tilde{\theta}_s. 
\end{align}

\subsection{Applying the Lorentz reciprocal theorem}\label{app:lorentz-perp}
Applying the Lorentz reciprocal theorem to the surface integral over the particle surfaces $\Sigma_{P_{1,2}^{\mbox{\tiny I,II}}}$ in Eq. \eqref{eqn:perp-corrDrag} for $(j,k) = (1,0)$, we obtain 
\begin{align}
\begin{split}
 \sum_{i=1,2}\iint_{\Sigma^{(0)}_{P_{i}^{\mbox{\tiny I,II}}}}\bm{\sigma}_\perp^{(1,0)} \cdot \tilde{\mathbf{n}}^{(0,0)} \cdot \mathbf{u}_{\perp}^\infty\mbox{ d}\Sigma = &   \iint_{\Sigma^{(0)}_I} [\bm{\sigma}_\perp^{(0,0)}] \cdot (-\hat{\mathbf{e}}_z) \cdot  \mathbf{u}_{\perp,D}^{(1,0)}  \mbox{ d}\Sigma \\
& -\iint_{\Sigma^{(0)}_I} [\bm{\sigma}_\perp^{(1,0)}]  \cdot (-\hat{\mathbf{e}}_z) \cdot \mathbf{u}^{(0,0)}_{\perp,D} \mbox{ d}\Sigma,
\end{split} \label{eqn:perp-reciprocalEqn1}
\end{align}
where 
\begin{align}
    & \mathbf{u}_{\perp,D}^{(1,0)} \cdot \hat{\mathbf{e}}_z = -\mathbf{u}_{\perp,D}^{(0,0)}\cdot \hat{\mathbf{n}}^{(1,0)} - \frac{\partial \mathbf{u}_{\perp,D}^{(0,0)}}{\partial z} h_\perp^{(1,0)}\cdot \hat{\mathbf{e}}_z ,\label{eqn:perp-reciprocal-integrand1}\\
  &    \hat{\mathbf{t}}^{(0,0)} \cdot  [\bm{\sigma}_\perp^{(1,0)}] \cdot \hat{\mathbf{e}}_z =  - \hat{\mathbf{t}}^{(0,0)} \cdot  \frac{\partial [\bm{\sigma}_\perp^{(0,0)}]}{\partial z }h_\perp^{(1,0)} \cdot \hat{\mathbf{e}}_z - \hat{\mathbf{t}}^{(0,0)} \cdot [\bm{\sigma}_\perp^{(0,0)}] \cdot \hat{\mathbf{n}}^{(1,0)} - \hat{\mathbf{t}}^{(1,0)} \cdot [\bm{\sigma}_\perp^{(0,0)}] \cdot \hat{\mathbf{e}}_z,\label{eqn:perp-reciprocal-integrand2}
\end{align}
for $\mathbf{x} \in \Sigma^{(0)}_I$.
The $\mathcal{O}(1)$ disturbance field is given by
\begin{align}
  \mathbf{u}^{(0,0)}_{\perp,D} =  \mathbf{u}_\perp^{(0,0)} - \hat{\mathbf{e}}_y = u_{\perp x }^{(0,0)} \hat{\mathbf{e}}_x + (u_{\perp y}^{(0,0)} -1) \hat{\mathbf{e}}_y,\label{eqn:perp-reciprocal-vector4}
\end{align}
and  the $\mathcal{O}(1)$ and $\mathcal{O}(\mbox{Ca})$ unit normal and tangential vectors to the fluid interface are 
\begin{align}
& \hat{\mathbf{t}}^{(0,0)} = \hat{\mathbf{e}}_x \mbox{ or }\hat{\mathbf{e}}_y, \label{eqn:perp-reciprocal-vector1}\\
  &   \hat{\mathbf{n}}^{(1,0)} = -\frac{\partial h_\perp^{(1,0)}}{\partial x }\hat{\mathbf{e}}_x  - \frac{\partial h_\perp^{(1,0)}}{\partial y} \hat{\mathbf{e}}_y,\label{eqn:perp-reciprocal-vector2}\\
    & \hat{\mathbf{t}}^{(1,0)} = \frac{\partial h_\perp^{(1,0)}}{\partial x} \hat{\mathbf{e}}_z \mbox{ or } \frac{\partial h_\perp^{(1,0)}}{\partial y} \hat{\mathbf{e}}_z.\label{eqn:perp-reciprocal-vector3}
\end{align}

Substituting Eqs. \eqref{eqn:perp-reciprocal-vector4}
- \eqref{eqn:perp-reciprocal-vector3} into Eqs. \eqref{eqn:perp-reciprocal-integrand1} and \eqref{eqn:perp-reciprocal-integrand2} yields
\begin{align}
    [\bm{\sigma}_\perp^{(0,0)}] \cdot (-\hat{\mathbf{e}}_z) \cdot  \mathbf{u}_{\perp,D}^{(1,0)}  = & 
     [\sigma_{\perp zz}^{(0,0)}] \left( \mathbf{u}_{\perp,D}^{(0,0)}\cdot \hat{\mathbf{n}}^{(1,0)} + \frac{\partial \mathbf{u}_{\perp,D}^{(0,0)}}{\partial z} h_\perp^{(1,0)}\cdot \hat{\mathbf{e}}_z  \right)   \\
    = &  [\sigma^{(0,0)}_{\perp zz}] \left(  - u^{(0,0)}_{\perp x} \frac{\partial h_\perp^{(1,0)}}{\partial x} - (u^{(0,0)}_{\perp y} -1) \frac{\partial h_\perp^{(1,0)}}{\partial y}  +  \frac{\partial u^{(0,0)}_{\perp z}}{\partial z} h^{(1,0)}  \right), 
\end{align}
and 
{
\begin{align}
   &  [\bm{\sigma}_\perp^{(1,0)} ]  \cdot (-\hat{\mathbf{e}}_z )  \cdot \mathbf{u}_{\perp,D}^{(0,0)}  \\
   & =  (- [\sigma_{\perp zx}^{(1,0)}] \hat{\mathbf{e}}_x - [\sigma_{\perp zy}^{(1,0)}] \hat{\mathbf{e}}_y  - [\sigma_{\perp zz}^{(1,0)}] \hat{\mathbf{e}}_z  ) \cdot  (u_{\perp x}^{(0,0)} \hat{\mathbf{e}}_x + (u_{\perp  y}^{(0,0)} -1  )\hat{\mathbf{e}}_y)  \\
    = &  \left(  \frac{\partial [\sigma_{\perp zx}^{(0,0)}]}{\partial z} h_\perp^{(1,0)} - [\sigma^{(0,0)}_{\perp xx} ] \frac{\partial h_\perp^{(1,0)}}{\partial x} - [\sigma_{\perp xy}^{(0,0)}]\frac{\partial h_\perp^{(1,0)}}{\partial y }+ [\sigma_{\perp zz}^{(0,0)}] \frac{\partial h_\perp^{(1,0)}}{\partial x }\right) u_{\perp  x}^{(0,0)} \\
    & +\left(  \frac{\partial [\sigma_{\perp zy}^{(0,0)}]}{\partial z} h_\perp^{(1,0)} - [\sigma^{(0,0)}_{\perp yx} ] \frac{\partial h_\perp^{(1,0)}}{\partial x}- [\sigma^{(0)}_{\perp yy} ] \frac{\partial h_\perp^{(1,0)}}{\partial y} +  [\sigma_{\perp zz}^{(0,0)}] \frac{\partial h_\perp^{(1,0)}}{\partial y }\right) (u_{\perp y}^{(0,0)} -1). 
\end{align} } 
Similar relations hold for the $\mathcal{O}(\delta)$ variables ($(j,k) = (0,1)$). 
The interfacial deformations $h^{(1,0)}$ and  $h^{(0,1)}$ and their partial derivatives with respect to $x$ and $y$ are obtained via numerical calculations. 

\section{Pair interaction: parallel flow}
\subsection{Leading order velocity field}
\label{app:para-leading-sol}
The leading order velocity field $\mathbf{u}_{\parallel}^{(0,0)} = u_{\parallel \rho}^{(0,0)} \hat{\mathbf{e}}_\rho + u_{\parallel \varphi}^{(0,0)} \hat{\mathbf{e}}_\varphi + u_{\parallel y}^{(0,0)} \hat{\mathbf{e}}_y $ obtained from \cite{Stimson1926} is axisymmetric and can be expressed in terms of the Stokes stream function: 
\begin{align}
    u_{\parallel \rho}^{(0,0)} = \frac{1}{\rho}\frac{\partial\psi }{\partial y} , \quad u_{\parallel y}^{(0,0)} = 1-\frac{1}{\rho} \frac{\partial \psi}{\partial \rho} ,
\end{align}
with $x = \rho\cos\varphi, z = r\sin\varphi.$  The stream function $\psi$ is given in bicylindrical coordinates $(\eta,\xi,\varphi)$ by 
\begin{align}
    \psi = (\cosh\xi-\cos\eta)^{-3/2}\sum_{n=1}^\infty (A_n \cosh(n-1/2)\xi+C_n\cosh(n+3/2)\xi) (P_{n+1}(\cos\eta) - P_{n+1}(\cos\eta)),
\end{align}
where $$\rho = \frac{c\sin\eta}{\cosh\xi-\cos\eta}, \quad y = \frac{c\sinh\xi}{\cosh\xi-\cos\eta},$$ $c = \sqrt{(L/2)^2-1},$ and $P_n$ is the Legendre polynomial of order $n.$ The coefficients $A_n$ and $C_n$ are given by
\begin{align}
    A_n = & -(2n+3)k\frac{2(1-e^{-(2n+1)\tau_1})+(2n+1)(e^{2\tau_1}-1)}{2\sinh(2n+1)\tau_1 + (2n+1)\sinh 2\tau_1},\\
     C_n = & (2n-1)k\frac{2(1-e^{-(2n+1)\tau_1})+(2n+1)(1-e^{-2\tau_1})}{2\sinh(2n+1)\tau_1 + (2n+1)\sinh 2\tau_1},
\end{align}
where $\tau_1 = \mbox{arccosh}(L/2)$ and 
\begin{align}
    k = \frac{c^2(-1)n(n+1)}{\sqrt{2}(2n-1)(2n+1)(2n+3)}.
\end{align}
\subsection{Pressure recovery}\label{app:pressure-recovery}
In section \ref{subsect:parallel-flow}, to calculate the drag force, we need the leading order jump in normal stress across the fluid interface. The viscous stress jump can be calculated from the solution for $\mathbf{u}^{(0,0)}_\parallel$ given by Stimson and Jeffery \cite{Stimson1926}. They did not give the pressure which we need to complete the drag calculation. Noting that $p_\parallel^{(0,0)}$ is harmonic and that we only need $p_\parallel^{(0,0)}$ along the fluid interface. We can identify a PDE for this interfacial pressure by rearranging the pressure equation into a 2D Poisson equation along the interface as 
\begin{align}
    \frac{\partial^2 p_\parallel^{(0,0)} }{\partial x^2}  +\frac{\partial^2 p_\parallel^{(0,0)} }{\partial y^2} = -  \frac{\partial^2 p_\parallel^{(0,0)} }{\partial z^2} \Bigg\vert_{z = 0} ,\quad -\infty < x,y < \infty.\label{eqn:app-para-pressureEqn1b}
\end{align}
In cylindrical coordinates $(\rho,\varphi,y)$ with $ x = \rho \cos\varphi$ and $z = \rho\sin\varphi, $ the RHS term in Eq. \eqref{eqn:app-para-pressureEqn1b} can be rewritten as 
\begin{align}
-\frac{\partial^2 p^{(0,0)}_\parallel }{\partial z^2} \Bigg\vert_{z = 0} = &  -\left( \frac{\partial^2 \rho}{\partial z^2} \frac{\partial^2 p^{(0,0)}_\parallel}{\partial \rho^2} + \left(\frac{\partial \rho}{\partial z}\frac{\partial }{\partial \rho} \frac{\partial \rho}{\partial z} + \frac{\partial\varphi}{\partial z} \frac{\partial }{\partial \varphi } \right) \frac{\partial p_\parallel^{(0,0)}}{\partial \rho}\right)\Bigg\vert_{\varphi = 0,\pi} \\
= & - \left(\sin^2\varphi \frac{\partial^2 p^{(0,0)}_\parallel}{\partial \rho^2} + \frac{\cos^2\varphi}{\rho} \frac{\partial p^{(0,0)}_\parallel}{\partial \rho} \right)\Bigg\vert_{\varphi = 0,\pi} \\
 & =  - \frac{1}{\rho} \frac{\partial p^{(0,0)}_\parallel}{\partial \rho}\Bigg\vert_{\varphi = 0,\pi} = -\frac{1}{x}\frac{\partial p^{(0,0)}_\parallel}{\partial x},
\end{align}
where $\partial p^{(0,0)}_{\parallel}/\partial \varphi = 0$ due to axisymmetry. 
Then, Eq. \eqref{eqn:app-para-pressureEqn1b} becomes 
\begin{align}
\frac{\partial^2 \tilde{p}_\parallel}{\partial x^2} + \frac{\partial^2 \tilde{p}_\parallel}{\partial y} = -  \frac{1}{x} \frac{\partial
     \tilde{p}_\parallel}{\partial x }. \label{eqn:app-para-pressureEqn2}
\end{align}
Note that  $p^{(0,0)}_\parallel$ also satisfies the $\hat{\mathbf{e}}_\rho$ component of the momentum equation
\begin{align}
        \frac{\partial p^{(0,0)}_\parallel}{\partial \rho } =    \frac{\partial^2 u^{(0,0)}_{\parallel \rho} }{\partial\rho^2 } + \frac{1}{\rho}\frac{\partial u^{(0,0)}_{\parallel \rho} }{\partial \rho } + \frac{\partial^2 u^{(0,0)}_{\parallel \rho}}{\partial y^2} - \frac{u^{(0,0)}_{\parallel \rho}}{\rho^2}.\label{eqn:app-para-pressureEqn3}
\end{align}
At the fluid interface, Eq. \eqref{eqn:app-para-pressureEqn3} can be rewritten as 
\begin{align}
        \frac{\partial p^{(0,0)}_\parallel}{\partial x } =    \frac{\partial^2 u^{(0,0)}_{\parallel x} }{\partial x^2 } + \frac{1}{x}\frac{\partial u^{(0,0)}_{\parallel x} }{\partial x } + \frac{\partial^2 u^{(0,0)}_{\parallel x}}{\partial y^2} - \frac{u^{(0,0)}_{\parallel x}}{x^2},\label{eqn:app-para-pressureEqn4}
\end{align}
where $u^{(0,0)}_{\parallel x} = u^{(0,0)}_{\parallel \rho} \cos\varphi = \pm u^{(0,0)}_{\parallel \rho}$ at $z = 0. $ Substituting Eq. \eqref{eqn:app-para-pressureEqn4} into Eq. \eqref{eqn:app-para-pressureEqn2} yields 
\begin{align}
   \frac{\partial^2 \tilde{p}_\parallel}{\partial x^2} + \frac{\partial^2 \tilde{p}_\parallel}{\partial y^2} = -  \frac{1}{x}   \left(  \frac{\partial^2 u^{(0,0)}_{\parallel x} }{\partial x^2 } + \frac{1}{x}\frac{\partial u^{(0,0)}_{\parallel x} }{\partial x } + \frac{\partial^2 u^{(0,0)}_{\parallel x}}{\partial y^2} - \frac{u^{(0,0)}_{\parallel x}}{x^2}  \right). 
\end{align}

\section{Capillary force calculation: perpendicular flow}
\label{app:capillaryForce}
The capillary unit vectors at the TCLs on the surfaces of particle I and II can be written as
\begin{align}
\begin{split}
    \tilde{\mathbf{n}}_{C,\perp}^{\text{\scriptsize I,II}} = \tilde{\mathbf{t}}_{C,\perp}^{\text{\scriptsize I,II}}  \times \hat{\mathbf{n}}_{\perp} =     \pm & \left[
\frac{-1+\cos\sigma\cosh\tau_1}{\cos\sigma-\cosh\tau_1}\hat{\mathbf{e}}_x + \frac{\sin\sigma\sinh(\mp\tau_1)}{\cos\sigma-\cosh\tau_1} \hat{\mathbf{e}}_y  \right.\\
& \ \ \  \left.+ \frac{-\cos\sigma+\cosh\tau_1}{c}\left(\mbox{Ca}\frac{\partial h^{(1,0)}}{\partial \tau} + \delta\frac{\partial h^{(0,1)}}{\partial \tau} \right) \hat{\mathbf{e}}_z 
    \right] ,
    \end{split}
\end{align}
where  
\begin{align}
\begin{split}
    \tilde{\mathbf{t}}_{C,\perp}^{\mbox{\scriptsize I,II}} = \mp
    \frac{\partial \mathbf{r}_{C,\perp}^{\text{\scriptsize I,II}}/\partial\sigma}{ \left\vert \partial\mathbf{r}_{C,\perp}^{\mbox{\scriptsize I,II}}/\partial\sigma \right\vert }
   = & \pm\left[ \frac{\sin\sigma\sinh\tau_1}{\cos\sigma-\cosh\tau_1} \hat{\mathbf{e}}_x + \frac{1-\cos\sigma\cosh\tau_1}{\cos\sigma-\cosh\tau_1}\hat{\mathbf{e}}_y  \right.
   \\
   & \left. +\frac{\cos\sigma-\cosh\tau_1}{c}\left( \mbox{Ca}\frac{\partial h^{(1,0)} }{\partial\sigma} (\mp\tau_1,\sigma) + \delta \frac{\partial h^{(0,1)}}{\partial \sigma }(\mp\tau_1,\sigma) \right)  \right] 
   \end{split}
\end{align}
are the unit tangent vectors to the TCLs, and 
\begin{align}
\begin{split}
    \hat{\mathbf{n}}^{\text{\scriptsize I,II}}_\perp = \hat{\mathbf{e}}_z +   \frac{1}{c} & \left(  (-1+\cos\sigma\cosh\tau_1 )  \left( \mbox{Ca}\frac{\partial h^{(1,0)}}{\partial \tau}   + \delta \frac{\partial h^{(0,1)}}{\partial \tau} \right) \right.\\
     &\left.+\sin\sigma\sinh(\mp\tau_1) \left( \mbox{Ca}\frac{\partial h^{(1,0)}}{\partial \sigma}   + \delta \frac{\partial h^{(0,1)}}{\partial \sigma} \right) \right) \hat{\mathbf{e}}_x \\
     + \frac{1}{c}  & \left(   (1-\cos\sigma\cosh\tau_1 )  \left( \mbox{Ca}\frac{\partial h^{(1,0)}}{\partial \sigma}   + \delta \frac{\partial h^{(0,1)}}{\partial \sigma} \right) \right.\\
    & \left.+\sin\sigma\sinh(\mp\tau_1) \left( \mbox{Ca}\frac{\partial h^{(1,0)}}{\partial \tau}   + \delta \frac{\partial h^{(0,1)}}{\partial \tau} \right) \right) \hat{\mathbf{e}}_y
    \end{split}
\end{align}
are  the unit normal vectors to the interface (the expansions are truncated after the $\mathcal{O}(\mbox{Ca})$ and  $\mathcal{O}(\delta)$ terms ).
Then, the capillary forces are computed to be 
\begin{align}
\begin{split}
    \mathbf{F}_C^{\text{\scriptsize I,II}} = &   \int_{-\pi}^\pi    \tilde{\mathbf{n}}_{C,\perp}^{\scriptsize \text{I,II}}  \frac{c \mbox{ d}\sigma}{\cosh\tau_1 - \cos\sigma} \\
    = & \pm  \int_{-\pi}^\pi \mbox{Ca} \frac{\partial h_\perp^{(1,0)}}{\partial\tau} (\mp\tau_1,\sigma) + \delta \frac{\partial h^{(0,1)}_\perp}{\partial\tau} (\mp\tau_1,\sigma) \mbox{ d}\sigma \hat{\mathbf{e}}_z,
    \end{split}\label{eqn:app-F_C}
\end{align}
and clearly, $\mathbf{F}_C^{\text{\scriptsize I}} = \mathbf{F}_C^{\text{\scriptsize II}} = \mathbf{F}_C$.

\nocite{*}

\bibliography{references}

\end{document}